\documentclass[usenatbib]{mn2e}
\usepackage{natbib,graphicx,times}
\usepackage{changebar}
\nochangebars

\newif\ifpdf 
  \ifx\pdfoutput\undefined 
    \pdffalse            
  \else 
    \pdfoutput=1         
    \pdftrue 
\fi

\ifpdf  
\usepackage[pdftex,colorlinks=true,
                   pdfstartview=FitV,
                   linkcolor=blue,
                   citecolor=red,
                   urlcolor=blue,
                   backref,pagebackref]{hyperref}
\fi

\title[Structure finding in photometric redshift galaxy surveys: An
extended friends-of-friends algorithm]{Finding structures in
  photometric redshift galaxy surveys: An extended
  friends-of-friends algorithm}

\author[C.~S.~ Botzler et al.]  
{C.~S.~Botzler,$^1$ 
  J.~Snigula,$^1$
  R.~Bender,$^{1,2}$
  U.~Hopp$^1$\\
  $^1$Universit\"ats-Sternwarte M\"unchen, Scheinerstr. 1, 
  D-81679 M\"unchen, Germany \\
  $^2$Max-Planck-Institut f\"ur extraterrestrische Physik,
  Giessenbachstr., D-85748 Garching, Germany}
 
\date{Accepted --; Received --; }

\pagerange{\pageref{firstpage}--\pageref{lastpage}} \pubyear{2002}

\begin{document}

\label{firstpage}

\maketitle

%
%

\begin{abstract}
  We present a modified version of the friends-of-friends (FOF)
  structure finding algorithm \citep{HG1982}, designed specifically to
  locate groups or clusters of galaxies in photometric redshift
  datasets. The main objective of this paper is to show that this
  extended friends-of-friends (hereafter EXT-FOF) algorithm yields
  results almost identical to the original FOF, if applied to a
  spectroscopic redshift dataset, and a rather conservative catalogue of
  structures, in case of a dataset with simulated photometric
  redshifts. Therefore, we create group catalogues for the first Center
  for Astrophysics Redshift Survey (CFA1; \citealt{Huchra1983}), as well
  as for the Las Campanas Redshift Survey (LCRS; \citealt{Shectman1996}),
  both of which being spectroscopic surveys, using FOF-algorithms. We
  then apply our new algorithm to said surveys and compare the
  resulting structure catalogue. Furthermore, we bestow simulated
  photometric redshifts on the LCRS galaxies, and use the EXT-FOF to
  detect structures, that we compare in size and composition to the
  ones found in the original, spectroscopic dataset. We will show that
  the properties of this modified algorithm are well understood and
  that it is suited for finding structures in photometric datasets.
  This is the first paper in a series of papers, dealing with the
  application of our new cluster finding algorithm to various
  photometric redshift galaxy surveys. 
\end{abstract}


%
%

\section{Introduction}
\label{s:introduction}

Optical and near-IR galaxy surveys, and especially the groups and
clusters of galaxies found therein, have been playing an important
role in our understanding of structure formation and cosmology. As has
been shown by \citet{PS1974}, \citet{Peebles1993}, and
\citet{Eke1996},
among others, the evolution of the mass-function of groups and
clusters of galaxies is highly sensitive to cosmological parameters,
the type of dark matter and the biasing of dark against baryonic
matter. Thus, comparison of the mass-function predicted by
semi-analytic structure formation models, like the Press-Schechter
formalism \citep{PS1974}, or N-body simulations with the observed
mass-function yields strong constraints on structure formation
scenarios and the cosmological paradigm \citep{BC1992, Ueda1993,
  JF1994, Bahcall1997, Bode2001}. Finding large-scale
structures in galaxy surveys is also an important step in examining
the evolution of galactic properties, like the luminosity function, in
high- and low-density environments 
(\citealt{Trentham1998, dePropris1999}; Drory et al.\ 2001a;
\citealt{Fried2001, Postman2001}).

With the development of the photometric redshift determination
techniques \citep{Baum62, Koo85, Brunner1999, FLY99, Benitez00,
  photred00_mod}, approximate redshifts became available for all
galaxies in a photometric multi-band survey, without having to do
time-consuming spectroscopic follow-up observations. This type of
galaxy survey is increasing in popularity. The list of surveys using
photometric redshift determination includes the Hubble Deep Fields
(HDF) North and South \citep{HDF96, HDFS00}, the Las Campanas Infrared
Survey (LCIR; \citealt{Marzke1999}, McCarthy et al.\ 2001a, b), the
Calar Alto Deep Imaging Survey (CADIS; Wolf et al.\ 2001a), 
the Classifying Objects by Medium-Band Observations Survey (COMBO-17;
Wolf et al.\ 2001b), 
the Munich Near-IR Cluster Survey (MUNICS; Drory
et al.\ 2001b), and the Fors Deep
Field (FDF; \citealt{FDF1}). However, photometrically
determined redshifts possess larger errors than spectroscopically
determined ones. The accuracy of the photometric redshift
determination is typically worse than the spectroscopic one by more
than two orders of magnitude. Thus, the photometric redshift errors
correspond to roughly 50 times the typical velocity dispersion of
bound structures, like galaxy groups or clusters.

The actual identification of groups or clusters of galaxies, either in
an observed or simulated catalogue, is a non-trivial enterprise. A
reliable detection of large-scale structures in an observed galaxy
catalogue is even more complicated if the galaxies have only
photometric redshifts or, even worse, none at all. Up to now, many
different techniques have been created for the purpose of structure
finding. All of them have their advantages and disadvantages and are
constructed to fit the specific features of their surveys, like the
wavelength of the observed radiation, or whether redshifts were
determined for all objects, and if so, what type of redshift
determination was used.  Giving a complete list of all types of
structure finding techniques applied to optical and near-IR data, and
their individual implementations would exceed the scope of this paper.
However, we give a brief overview over the more important ones and
their working principles, to better place our work in the context of
structure finding.

The pioneering work in this field was achieved by \citet{Abell1958},
followed by \citet{ACO1989}, \citet{LP1996}, and \citet{Dalton1997},
amongst others. Their counts-in-cells method looks for surface density
enhancements in circular apertures of fixed physical radius. A
modification to this approach was used by \citet{Zwicky6168},
\citet{TG1976}, \citet{Couch1991}, and \citet{Plionis1991}, who
searched for surface density enhancements by method of
isopleth-contours. Another rather famous method is the hierarchical
clustering technique \citep{Materne1978, T1980, hopp1985, Tully1987,
  Gourgoulhon1992}. It identifies structures on the basis of
optimising typical cluster or group properties, such as spatial
separation or luminosity density, by combining individual galaxies
into groups, with the advantage of delivering a list of local
concentrations within the groups, and superstructures at given values
of the property that is to be optimised. One of the more recently
developed methods is the matched or generalised likelihood filter
\citep{Postman1996, SB1998, Kepner1999, Lobo2000}.  On the basis of
Bayesian probability theory, the galaxy catalogue is convolved with a
set of models, the filter, describing typical cluster properties, like
the radial profile or the luminosity function, thus determining the
likelihood of having clusters with given positions and richnesses in
the input dataset.  Voronoi tessellation has also been used as a means
of detecting large-scale structures \citep{Ramella2001, Kim2002}. In
this case, a unique plane or volume partition of the galaxy
distribution is done, yielding a measure for the local densities, thus
enabling identification of groups or clusters as significantly
over-dense regions. Techniques looking for spherical density
enhancements have been used by \citet{LC1994} and \citet{CL1996}. Some
more exceptional approaches are the wavelet decomposition
\citep{SBM1990, PJM1995}, the HOP algorithm \citep{EH1998},
the cluster red sequence approach \citep{GY2000},
or the cut-and-enhance method \citep{Goto2002}.

Probably the most extensively used structure finding algorithm is the
friends-of-friends technique or percolation algorithm. This method
looks for number density enhancements in three dimensions by searching
for galaxy pairs, that are closer to one another than a given cut-off
separation.  
\cbstart 
Consequently, the friends-of-friends does not
suffer a bias as far as the radial profile, or luminosity function are
concerned, as is for example the case for the matched filter
technique.  
\cbend 
Up to now, the friends-of-friends (FOF) approach
has only been applied to spectroscopic redshift surveys \citep{HG1982,
  GH1983, RGH1989, RPG1997, TB1998, GMCP2000, MML2000, Tucker2000,
  RGPdC2002}, or N-body simulations \citep{DEFW1985, EFWD1988, LC1994,
  CL1996, VLS2000}. In the first case, the spectroscopic redshift is a
combination of the Hubble expansion and the peculiar velocity of the
galaxy, thus being a very good estimate of the true distance to the
galaxy. In the second case, the exact position in three-dimensional
space is known. Applying this algorithm to a galaxy catalogue with
photometric redshifts constitutes a problem, due to the less than
perfect distance information. If the redshift errors are not taken
into account, the resulting structure catalogue will contain highly
unphysical groups and clusters. If the errors are taken into account,
yet the algorithm itself is not changed, the procedure yields
structures, that are extremely elongated in redshift $\left(\Delta z
  \sim 1 \right)$. Thus it proves necessary to develop a new type of
FOF algorithm, the extended friends-of-friends, for photometric
redshift datasets, that includes the redshift error information and at
the same time cuts down on unrealistically elongated structures.

Section \ref{s:fofalgs} gives a brief overview of the original FOF
technique of \citet{HG1982} and introduces the EXT-FOF algorithm. The
following two sections prove, that in case of galaxy catalogues with
spectroscopically determined redshifts, both algorithms yield almost
identical group catalogues. In Section \ref{s:cfa}, both
friends-of-friends algorithms are applied to the spectroscopic CFA1
survey, following the recipe given by \citet{GH1983}. The resulting
group catalogues are compared on an object-to-object basis and the
deviations are explained. The same is done for the LCRS in Section
\ref{s:lcrs}, following the approach of \citet{Tucker2000}. In order
to show that our EXT-FOF technique provides a conservative group and
cluster catalogue for photometric galaxy surveys, we simulate
photometric redshifts for the LCRS galaxies in Section
\ref{s:lcrsphotred}. We then apply our EXT-FOF to this new dataset and
compare the detected structures to the ones found with the EXT-FOF
algorithm in the original, spectroscopic dataset.

This paper is the first in a series, dealing with the search for
large-scale structures, like groups and clusters of galaxies, in
photometric redshift datasets. In the next paper, we will apply our
new cluster finding technique to MUNICS and test the plausibility of
the found structures, using colour-magnitude diagrams, Voronoi
tessellations, and a likelihood approach. The same will be
performed for the FDF galaxy catalogue in our third paper.

%
%

\section{Friends-of-friends (FOF) algorithms}
\label{s:fofalgs}

The friends-of-friends algorithm, that was invented by \citet{HG1982},
is one of the most frequently used cluster finding techniques. It was
designed to find number overdensities in spectroscopic galaxy surveys
and has also been modified to look for structures in simulated galaxy
datasets. The FOF approach is very straightforward and has no need for
complicated cluster models. It looks for galaxy pairs, that are closer
to one another than a given cut-off separation.  One of the resulting
advantages is its insensitivity to cluster shape.  However, like most
of the other structure finding techniques, it has the drawback of
being rather slow in its application. Changing the cut-off parameters
requires a new run of the procedure, which makes the optimisation of
the cut-off values time-consuming.

\subsection{The Huchra \& Geller friends-of-friends algorithm}
\label{ss:fof}

The original FOF algorithm was created to look for groups and clusters
of galaxies in the magnitude-limited CFA1 redshift survey
\citep{HG1982, Huchra1983, GH1983}. It makes use of three basic pieces
of information, the positions of the galaxies in right ascension and
declination, and their spectroscopic redshifts.

\begin{figure}
  \includegraphics[width=\columnwidth]{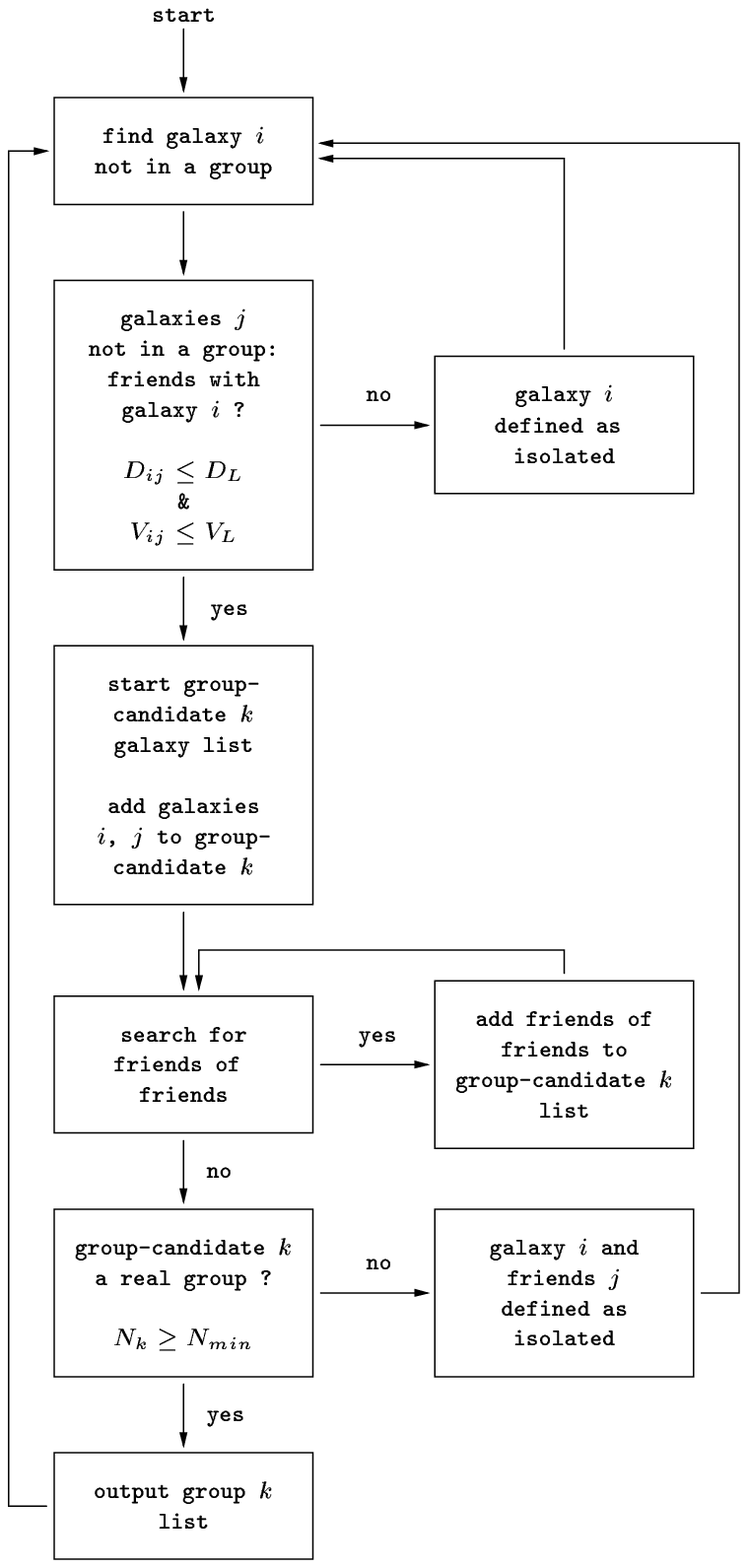}
  \caption{Flow chart for the Huchra \& Geller friends-of-friends algorithm}
  \label{fofflow}
\end{figure}

A slightly modified version of the algorithm's flow chart given by
\citet{HG1982} is shown in Fig.\ \ref{fofflow}. First, an object $i$
from the galaxy catalogue, that has not yet been assigned to any group,
is chosen. Then friends $j$ of that galaxy are searched for. They have
to fulfill two criteria. Their projected separation $D_{ij}$ from the
first galaxy has to be less than a critical value $D_L$:
\begin{equation}
\label{Dfof}
D_{ij} := 2 \sin \frac{\theta_{ij}}{2} \frac{\overline{V}}{H_0} 
\leq D_L ,
\end{equation}
where $\theta_{ij}$ is the angular separation between galaxies $i$ and
$j$, $H_0$ is the Hubble constant, and $\overline{V}$ is the mean
velocity of the galaxy pair:
\begin{equation}
\label{overV}
\overline{V} = \frac{v_i + v_j}{2},
\end{equation}
$v_i$ and $v_j$ being the velocities of the individual galaxies.
Furthermore, their separation $V_{ij}$ in velocity-space has to be
less than a second critical value $V_L$:
\begin{equation}
\label{Vfof}
V_{ij} := \left|v_i - v_j\right| 
\leq V_L .
\end{equation}
If there are no friends for galaxy $i$, it is called an isolated
object and removed from the catalogue of possible cluster members. If
friends are found for galaxy $i$, a list for group-candidate $k$ is
initiated, containing galaxy $i$ and all of its friends $j$, that
fulfill eqs.\ (\ref{Dfof}) and (\ref{Vfof}). The search for friends is
extended to the surroundings of the galaxies $j$. All friends found
are once more added to the group-candidate $k$. This is done until no
further friends of friends can be detected. The group-candidate $k$ is
called a real group, if the number $N_{k}$ of its members exceeds the
limit $N_{min}$:
\begin{equation}
\label{Nminfof}
N_k \geq N_{min} .
\end{equation}
If this is not the case, all members of said group-candidate are
removed from the catalogue. The next galaxy from the catalogue is then
taken and its surroundings are searched for friends.

As can be seen easily, this algorithm is commutative and yields
reproducible results, which are very important features for any
structure finding technique. 

The choice for the linking parameters $D_L$ and $V_L$ depends on the
properties of the input galaxy dataset. In principle, both parameters
can be chosen to remain either fixed or to vary with redshift. The
minimum group size limit $N_{min}$ is usually set to $3$. Thus only
galaxy pairs are excluded from the final group catalogue. 

\subsection{The extended friends-of-friends algorithm}
\label{ss:extfof}

\begin{figure*}
  \includegraphics[width=\textwidth]{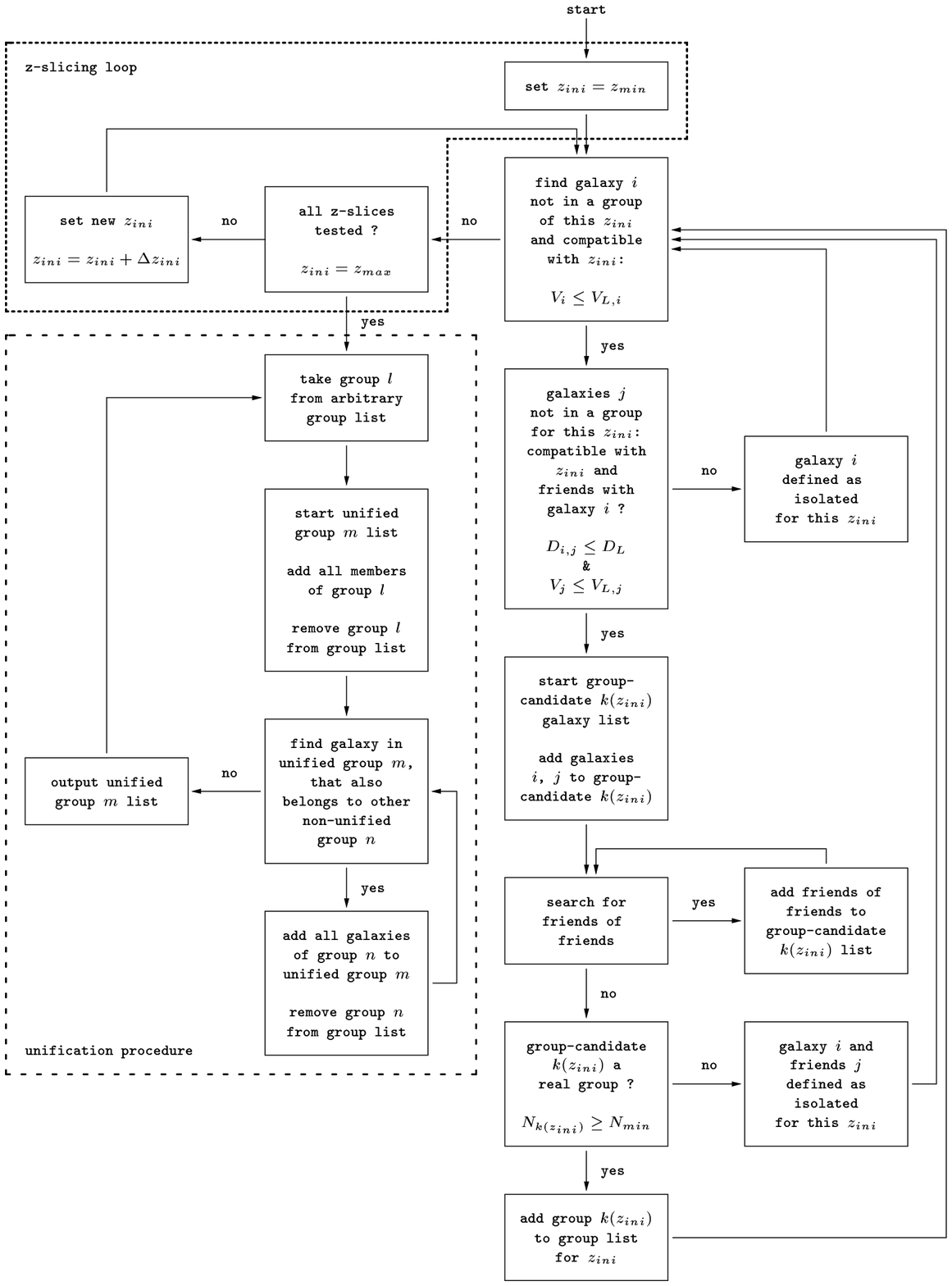}
  \caption{Flow chart for the extended friends-of-friends algorithm}
  \label{extfofflow}
\end{figure*}

Due to the relatively large errors in the photometric redshifts, the
friends-of-friends algorithm cannot be used for this type of dataset
without some modifications. If the photometric redshift errors are not
taken into account, a large fraction of the resulting groups and
clusters are not physically bound objects.  While the linking velocity
$V_L$ of the original FOF usually corresponds roughly to typical
cluster velocity dispersions, photometric redshift errors are normally
larger by a factor of 50.  This implies that the deviation between the
measured and the true distance to a galaxy might be off by about 50
times the linking velocity. Identification of these galaxies as group
members is impossible with the original FOF technique. On the other
hand, galaxies, whose true positions on the velocity axis are deviant
by about 100 $V_L$, could be combined into groups. Simply including
the redshift errors of the individual galaxies into the linking
criteria does not solve this dilemma either. The resulting structures
would get unreasonably extended in redshift. Allowing for the
necessary large value in the linking velocity would enable huge chains
of galaxies to be linked together by the FOF algorithm. Thus, a
modified version of the FOF algorithm, the extended friends-of-friends
or EXT-FOF, becomes necessary for photometric redshift galaxy surveys.

The EXT-FOF technique utilises the same informations as the original
FOF, i.e.\ right ascensions, declinations and, in this case,
photometric redshifts of the galaxies. Furthermore the individual
redshift error of each galaxy is taken into account.

The algorithm itself consists of three parts: In its inner loop, the
technique is almost identical to the Huchra \& Geller method, with
slightly altered linking conditions. The search for groups is done in
various a priori redshift-slices, meaning only galaxies that are
compatible with a given value of redshift are taken into account for
cluster finding. This redshift-slicing constitutes the outer loop. As
a result, there is a catalogue of structures for every redshift-slice.
Every galaxy can only be a member of one structure of the catalogue
belonging to a given $z$-slice. Yet it can also belong to other
structures in the other $z$-slices. The third part of the algorithm is
the unification of all structures that have at least one member in
common. This guarantees that in the final catalogue every galaxy can
only belong to one group or cluster.

The underlying idea for the modification of the linking conditions is
the following: The a priori redshift slices $z_{ini}$ approximate
roughly the mean redshifts $\overline{V}$ of the original
FOF-technique. So eq.\ (\ref{Dfof}) changes into
\begin{equation}
\label{Dextfof}
D_{ij} := 2 \sin \frac{\theta_{ij}}{2} D \left(z_{ini}\right) 
\leq D_L .
\end{equation}
$D \left(z_{ini}\right)$ is the distance to $z_{ini}$ in Mpc, being
either 
\begin{equation}
\label{Dlowz}
D \left(z_{ini}\right) := \frac{c z_{ini}}{H_0}
\end{equation}
in the case of a low-redshift approximation, or more correctly 
\begin{equation}
\label{Dhighz}
D \left(z_{ini}\right) := 
d_A \left(H_0, \Omega_M, \Omega_\Lambda, z_{ini}\right)
\end{equation} 
in the case of a cosmologically exact distance measurement.
$d_A\left(H_0, \Omega_M, \Omega_\Lambda, z_{ini}\right)$ is the
angular distance to $z_{ini}$ for a cosmology with given $H_0$,
$\Omega_M$ and $\Omega_\Lambda$ \citep{CPT:1992}. Eq.\ (\ref{Vfof})
translates into two equations, one for each of the galaxies $i$ and
$j$:
\begin{equation}
\label{Vextfofi}
V_{i} := \left|v_i - c z_{ini}\right| 
\leq \frac{V_L}{2} ,
\end{equation}
\begin{equation}
\label{Vextfofj}
V_{j} := \left|v_j - c z_{ini}\right| 
\leq \frac{V_L}{2} .
\end{equation}
If the individual redshift errors $\delta z_i$ and $\delta z_j$ of the
galaxies $i$ and $j$ are taken into account, the left side of
relations (\ref{Vextfofi}) and (\ref{Vextfofj}) changes:
\begin{equation}
\label{Vextfoferri}
V_{i} \leq \left[\left(V_L / 2\right)^2 + \left(c \delta
    z_i\right)^2\right]^{1/2} =: V_{L, i},
\end{equation}
\begin{equation}
\label{Vextfoferrj}
V_{j} \leq \left[\left(V_L / 2\right)^2 + \left(c \delta
    z_j\right)^2\right]^{1/2} =: V_{L, j}.
\end{equation}
Obviously, in the case of very small redshift errors, eqs.\ 
(\ref{Vextfofi}) and (\ref{Vextfofj}) are good approximations for
eqs.\ (\ref{Vextfoferri}) and (\ref{Vextfoferrj}).

A flow chart for the EXT-FOF algorithm is shown in Fig.\ 
\ref{extfofflow}. First, the minimal redshift $z_{ini} = z_{min}$ is
chosen for structure finding. A catalogue of groups, that belong to
this $z_{ini}$-slice, is created as follows: An object $i$, that has
not yet been assigned to any group belonging to this redshift-slice,
is chosen from the catalogue. Unlike the original FOF, this galaxy
also has to be compatible with the chosen $z_{ini}$, meaning it has to
fulfill eq.\ (\ref{Vextfoferri}) (or eq.\ (\ref{Vextfofi})). Then,
friends $j$ of that galaxy are searched for. They also have to be
compatible with the chosen $z_{ini}$, i.e.\ they have to satisfy eq.
(\ref{Vextfoferrj}) (or eq.\ (\ref{Vextfofj})) and have to be closer to
galaxy $i$ than the cut-off distance $D_{L}$ (eq.\ (\ref{Dextfof})). If
no friends can be found, object $i$ is moved to a list of isolated
galaxies. If friends can be found, a group-candidate $k
\left(z_{ini}\right)$ is initiated and the galaxies $i$ and $j$ are
added to it. The surroundings of the galaxies $j$ are searched for
companions fulfilling eqs.\ (\ref{Dextfof}) and (\ref{Vextfoferrj})
(or (\ref{Vextfofj})), and the loop is repeated until no further
friends can be found. A group-candidate is called a real group, if the
number $N_{k \left(z_{ini}\right)}$ of galaxies belonging to it
satisfies the relation
\begin{equation}
\label{Nminextfof}
N_{k \left(z_{ini}\right)} \geq N_{min} 
\end{equation}
and the group is then added to the catalogue of structures for the
given redshift-slice. If this is not the case, the galaxy $i$ and its
friends are moved to the list of isolated objects. The next galaxy
from the catalogue is then taken and its surroundings are searched for
friends. If all galaxies are assigned either to a group or to the
isolated list and no galaxies are left in the input catalogue, the
search is continued in the next redshift-slice. For this, $z_{ini}$ is
increased by a value $\Delta z_{ini}$ and the original input catalogue
of galaxies is restored. The search for friends is then repeated as
described above. Once the value of $z_{ini}$ has reached the limit
$z_{max}$, the search is stopped. So far, this technique yields
$\left(z_{max} - z_{min}\right) / \Delta z_{ini}$ individual
catalogues of clusters.  Since one galaxy can belong to different
clusters in different redshift-slices, a unification of these groups
is necessary to remove this ambiguity. To do this, a group $l$ of any
arbitrary $z_{ini}$-slice is taken. A, so far empty, unified group $m$
is created. All galaxies of the above mentioned group $l$ are added to
the unified group $m$ and group $l$ is removed from the group
catalogue.  Next, the unification procedure looks for a galaxy in
unified group $m$ that also belongs to another, non-unified, group
$n$, irrespective of the redshift-slice that this group belongs to. If
no such galaxy can be found, unified group $m$ is complete, and the
process is repeated for another group $l$. If there is a galaxy in $m$
that belongs to a group $n$, as well, all galaxies of group $n$ are
added to the unified group $m$ and $n$ gets removed from the
catalogue. This loop is repeated until unified group $m$ is complete
and none of its members are contained in any of the non-unified groups
any more. The result is a catalogue of disjoint structures.

Like Huchra \& Geller's version, this new algorithm is commutative and
yields reproducible results, i.e.\ the sequence of the input galaxies
does not play a role.  

One of this technique's objectives is to reproduce group catalogues
found with the Huchra \& Geller FOF in the case of spectroscopic
datasets. To do this, the values of $z_{ini}$ have to approximate
every possible value of $\overline V$ for every galaxy pair.
Theoretically, this can be reached by using a continuum of
$z_{ini}$-values, ranging from the minimum $\overline V$ of the two
galaxies in the dataset that have the lowest redshifts to the maximum
$\overline V$ of the galaxy pair with highest redshifts. Of course
this is not feasible in practice. Instead, a discrete set of $z_{ini}$
values is used with a very small spacing $\Delta z_{ini}$. The
boundaries $z_{min}$ and $z_{max}$ are simply set to the minimum and
maximum redshifts of the galaxies used for structure finding.

There are two basic differences between the two FOF techniques: While
the Huchra \& Geller technique tests, whether eq.\ (\ref{Dfof}) is
fulfilled for a galaxy pair at only one redshift, the new algorithm
tests, whether eq.\ (\ref{Dextfof}) is satisfied for said galaxy pair
at a multitude of redshifts. Furthermore, with the original algorithm,
galaxies that are close to one another in projection can be linked
together, even though not all of them are close to one another in
redshift. Thus, chains of $N \geq N_{min}$ galaxies can be linked
together into groups, that are very elongated in the redshift. In the
case of the extended FOF, all $N \geq N_{min}$ galaxies have to be
compatible with a given redshift $z_{ini}$ in order to be called a
group. This makes the above mentioned outcome of elongated galaxy
chains highly unlikely and is the reason why the new algorithm is
well-suited for cluster finding in photometric redshift surveys. The
next Sections (\ref{s:cfa}, \ref{s:lcrs}, and \ref{s:lcrsphotred})
show all of the possible effects resulting from these discrepancies
and prove the validity of the extended friends-of-friends for
photometric redshift datasets.

\section{Application to the CFA1 redshift survey}
\label{s:cfa}

\subsection{The galaxy catalogue}
\label{ss:cfasurvey}

The CFA1 redshift survey \citep{Huchra1983} is a magnitude limited,
spectroscopic galaxy survey, containing all the galaxies of the Zwicky
\citep{Zwicky6168, Zwicky1971} or \citet{Nilson1973} catalogues, that
satisfy the following selection criteria:
\begin{equation}
m_{pg} \leq 14.5 \; \textrm{mag ,}
\end{equation}
and 
\begin{equation}
b \geq 40\degr , \; \delta \geq 0\degr , 
\quad \textrm{or} \quad
b \leq -30\degr , \; \delta \geq -2\fdg5 .
\end{equation}
The magnitudes are given in the B(0)-Zwicky system. The original
catalogue covers 2401 galaxies.

In this paper, an electronic version of the CFA1 galaxy catalogue,
created in May 1997, is used. This new catalogue contains 2396
galaxies with corrected magnitudes and redshifts.

\subsection{Creation of the group catalogues}
\label{ss:cfagroups}

Following the recipe of \citet{GH1983}, two group catalogues are
created, one by using the original FOF technique, the other by
applying the new EXT-FOF algorithm. A summary of the \citet{GH1983}
treatment of the galaxy data and choice of the linking parameters is
given below:

First, the heliocentric galaxy velocities are corrected for a dipole
virgocentric flow
\begin{equation}
V_V = V_{in} \left[\sin\delta_i \sin\delta_V + \cos\delta_i 
\cos\delta_V \cos\left(\alpha_i - \alpha_V\right)\right] ,
\end{equation}
where $V_{in}$ is the infall velocity, which is set to $300 \textrm{
  km s}^{-1}$. $\alpha_V$ and $\delta_V$ are the right ascension and
declination of the Virgo cluster ($\alpha_V = 12^{\mathrm{h}} 28\fm7$;
$\delta_V = 12\degr 19\farcm1$; J1950; \citealt{RPG1997}), and
$\alpha_i$, and $\delta_i$ are the right ascension and declination of
galaxy $i$.  Furthermore, a correction for the solar motion with
respect to the local group is made \citep{RPG1997}:
\begin{equation}
V_G = 300 \textrm{ km s}^{-1} \sin l_i \cos b_i .
\end{equation}
$l_i$ and $b_i$ are the galactic longitude and latitude of galaxy
$i$. All galaxies with corrected velocities less than $300 \textrm{ km
  s}^{-1}$ are given an indicative velocity of $300 \textrm{ km
  s}^{-1}$. This is done to avoid the singularity at $0 \textrm{ km
  s}^{-1}$.

Only galaxies with velocities less than $12000 \textrm{ km s}^{-1}$
are accepted for the cluster search, while the mean velocity of the
groups is limited to less than $8000 \textrm{ km s}^{-1}$.

A Hubble constant of $H_0 = 100 \textrm{ km s}^{-1}\textrm{ Mpc}^{-1}$
is used, to be in accordance with the approach used by \citet{GH1983}.

\begin{figure}
  \includegraphics[width=\columnwidth]{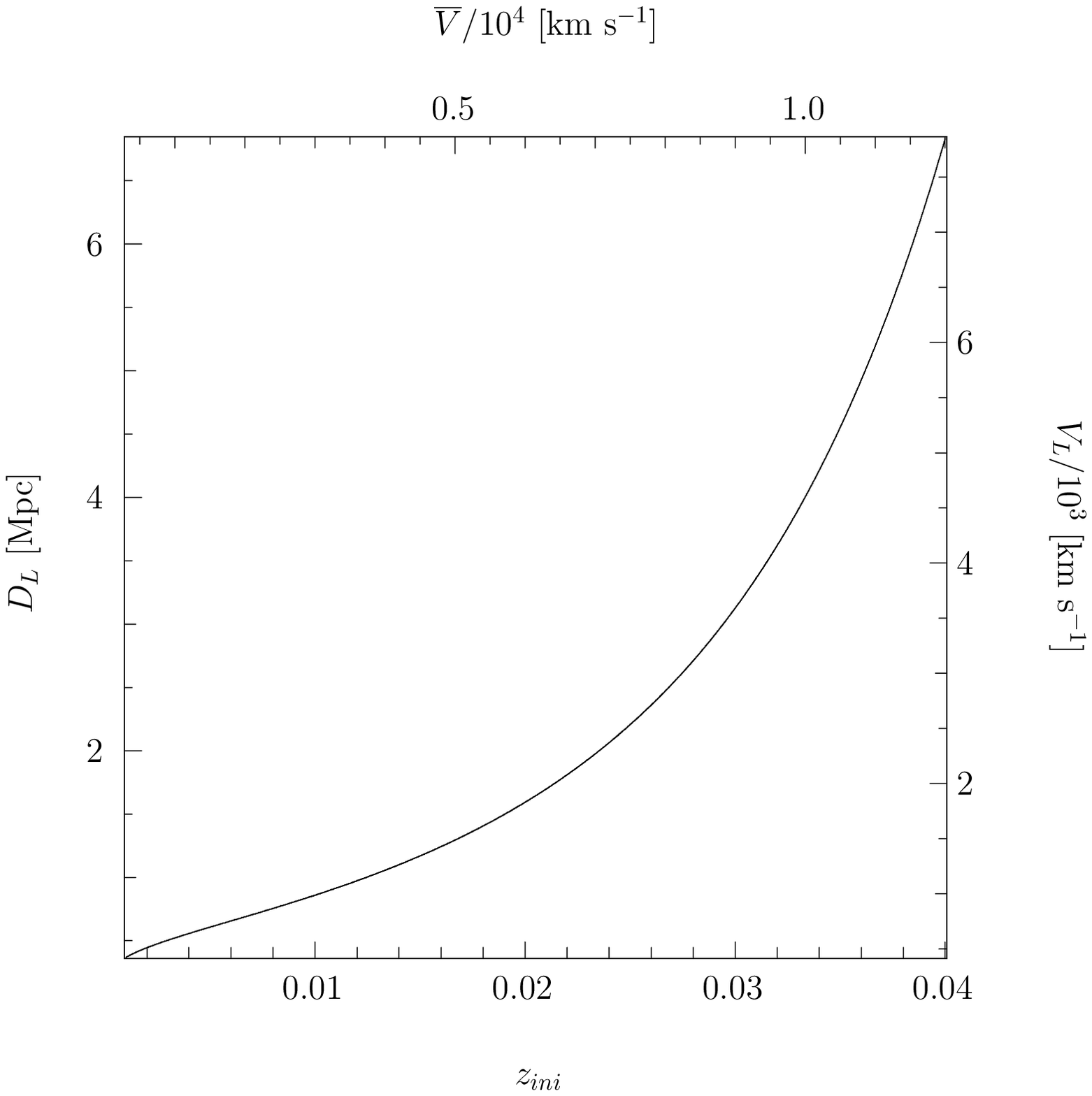}
  \caption{Variation of the linking criteria $D_L$ and $V_L$ for the
    CFA1 group catalogues, as a function of the FOF mean separation
    $\overline V$, or the EXT-FOF redshift-slicing $z_{ini}$}
  \label{dlvlcfa}
\end{figure}

The linking parameters are chosen to vary with redshift, in order to
compensate for the variation in the sampling of the luminosity
function,
\cbstart
as described in \citet{HG1982}, and \citet{GH1983}:
\cbend
\begin{equation}
\label{varDL}
D_L = D_0 R ,
\end{equation}
and
\begin{equation}
\label{varVL}
V_L = V_0 R ,
\end{equation}
with
\begin{equation}
R = \left[\int_{-\infty}^{M_{ij}} \Phi\left(M\right) dM
  \left(\int_{-\infty}^{M_{lim}} \Phi\left(M\right)
    dM\right)^{-1}\right]^{-1/3} .
\end{equation}
$\Phi\left(M\right)$ is the Schechter luminosity function with $\alpha
= -1.30$, $M^* = -19.40 \; \textrm{mag}$, and $\Phi^* = 0.0143 \;
\textrm{Mpc}^{-3}$.
\begin{equation}
M_{lim} = m_{lim} - 25 - 5 \log \left(V_{fid}/H_0\right) ,
\end{equation}
and
\begin{equation}
\label{Mijfof}
M_{ij} = m_{lim} - 25 - 5 \log \left(\overline{V}/H_0\right) ,
\end{equation}
in case of the original FOF technique. For cluster finding with the
EXT-FOF algorithm, eq.\ (\ref{Mijfof}) changes to
\begin{equation}
\label{Mijextfof}
M_{ij} = m_{lim} - 25 - 5 \log \left(c z_{ini}/H_0\right) .
\end{equation}
The fiducial velocity $V_{fid}$ is set to $V_{fid} = 1000 \textrm{ km
  s}^{-1}$, the projected separation $D_0$ and the velocity difference
$V_0$ at the fiducial velocity are chosen as $D_0 = 0.52 \textrm{ Mpc}$
and $V_0 = 600 \textrm{ km s}^{-1}$. The resulting variation of the
linking criteria $D_L$ and $V_L$ as a function of $\overline V$, or
$z_{ini}$ respectively, is shown in Fig.\ \ref{dlvlcfa}. 
\cbstart
The described variation of the linking parameters enables the
  algorithm to detect only structures that have a given minimal number
  overdensity relative to the mean, completely independent of the
  structure's redshift.
\cbend

The CFA1 group catalogue published by \citet{GH1983} contained only
groups with more than two members, so we set $N_{min} = 3$.

Since the CFA1 is a local galaxy survey, the low-redshift
approximation, eq.\ (\ref{Dlowz}), is used for the distance
calculation in the EXT-FOF. Because of the very small redshift errors
of this spectroscopic survey, eqs.\ (\ref{Vextfofi}) and
(\ref{Vextfofj}) are utilised as the velocity linking criteria. The
redshift-spacing is set to $\Delta z_{ini} = 10^{-5}$, to make sure
that every possible value of $\overline V$ can be approximated.

\subsection{Comparison of the group catalogues}
\label{ss:cfacomp}

\begin{table*}
  \caption{Comparison of the CFA1 FOF and EXT-FOF structures}
  \label{tablecfa}
  \begin{tabular}{lccccc}
    \hline
    & category & FOF & EXT-FOF & \% FOF & \% EXT-FOF \\
    \hline
    total &  & 176 & 165 & 100 & 100 \\
    &&&&& \\
    identical & 1 & 122 & 122 & 69.3 & 73.9 \\
    only FOF struct.\ & 2 & 4 & -- & 2.3 & -- \\
    only EXT-FOF struct.\ & 3 & -- & 8 & -- & 4.8 \\
    FOF struct.\ larger than EXT-FOF & 4 & 5 & 5 & 2.8 & 3.0 \\
    EXT-FOF struct.\ larger than FOF & 5 & 12 & 12 & 6.8 & 7.3 \\
    FOF struct.\ is combination of EXT-FOF structs.\ & 6 & 0 & 0 & 0.0 &
    0.0 \\
    EXT-FOF struct.\ is combination of FOF structs.\ & 7 & 28 & 13 & 15.9
    & 7.9 \\
    FOF and EXT-FOF structs.\ have some elements in common & 8 & 5 & 5 & 2.8 & 3.0 \\
    &&&&& \\
    FOF structs.\ found with EXT-FOF alg.\ &  & 172 & -- & 97.7 & -- \\
    \hline
  \end{tabular}

\end{table*}

The catalogue of structures, resulting from the application of the
original FOF algorithm, contains 176 groups and clusters, comprising a
total of 1480 galaxies. The composition of these FOF structures is
slightly deviant from the ones published by \citet{GH1983}. This minor
discrepancy is due to the different input galaxy catalogues.

Application of the EXT-FOF algorithm yields 165 groups, containing
1518 galaxies in total. Thus, in the case of the CFA1, the EXT-FOF
shows a tendency to identify somewhat larger structures.

In order to compare the algorithms, we do an object-to-object
comparison of the group members.  We define eight categories for the
level of agreement in the group composition:
\begin{enumerate}
  \renewcommand{\theenumi}{(\arabic{enumi})}
  \item The FOF and EXT-FOF group have identical composition
  \item The group is only found with FOF, i.e.\ no galaxy in the
    group is a member of an EXT-FOF group  
  \item The group is only found with EXT-FOF, i.e.\ no galaxy in the
    group is a member of a FOF group  
  \item The EXT-FOF group is a subset of the FOF group, i.e.\ the FOF
    group has more members than the EXT-FOF, and none of those surplus
    members belong to any other EXT-FOF group 
  \item The FOF group is a subset of the EXT-FOF group, i.e.\ the EXT-FOF
    group has more members than the FOF, and none of those surplus
    members belong to any other FOF group 
  \item The FOF group is a combination of multiple EXT-FOF groups, and
    can also contain further galaxies that are not part of any other
    EXT-FOF group 
  \item The EXT-FOF group is a combination of multiple FOF groups, and
    can also contain further galaxies that are not part of any other
    FOF group 
  \item The FOF and EXT-FOF group have some members in common, but do
    not fall under any of the above mentioned criteria
\end{enumerate}
This classification scheme facilitates the examination of the
intrinsic characteristics of the two friends-of-friends algorithms, as
can be seen in the next Section \ref{ss:cfaanalysis}.

Table \ref{tablecfa} shows the statistics of the comparison between
the FOF and EXT-FOF structures. The first column provides a short
description of the categories used. In the second column, the
corresponding categories are listed for easy reference. The third and
fourth column contain the number of FOF and EXT-FOF groups respectively,
that fall under the specified category, while the fifth and sixth
column show the percentage of these groups. 

122 groups or clusters are recovered identically (category 1) by both
algorithms, corresponding roughly to 72\% of the FOF or EXT-FOF
structures. 17 of the FOF groups are found by the EXT-FOF algorithm
either as slightly enlarged (category 5), or reduced (category 4)
structures. 28 FOF structures are recovered by the EXT-FOF as a
combination of multiple FOF groups (category 7).  Further five FOF
structures are found by the EXT-FOF, suffering from combinations of
the categories 4, 5, 6, and 7 (i.e.\ category 8). Only four of the 176
FOF groups are not found with the EXT-FOF algorithm (category 2).
Thus, a total of 172 FOF structures are recovered by the EXT-FOF
algorithm, corresponding to a recovery rate for the EXT-FOF algorithm
of almost 98\%. On the other hand, the EXT-FOF technique finds eight
additional groups (category 3), that are not part of the FOF group
catalogue, leading to a spurious detection rate for the EXT-FOF
technique of less than 5\%, provided that the FOF technique delivers a
complete structure catalogue. A closer look at the category 2 groups
shows, that they are extremely small. Three of them have only three
members and one of them contains four members.  Furthermore, they are
very elongated in the redshift direction, an attribute that might
speak against them as possible galaxy groups, anyway. The category 3
EXT-FOF groups are also very small. Four of them have three members,
the other four contain four members. All things considered, in the
case of the CFA1 group determination, the EXT-FOF algorithm yields
results very similar to the FOF technique. A slight tendency to find
larger groups with the EXT-FOF can be seen here. The cause for the
discrepancies between the two friends-of-friends techniques is
explained in Section \ref{ss:cfaanalysis}.

\subsection{Analysis of the discrepancies}
\label{ss:cfaanalysis}

The deviations between the two group catalogues can be ascribed to two
effects: Either the EXT-FOF algorithm is able to link two galaxies
together, that the original FOF cannot; Or the EXT-FOF algorithm can
not link $N_{min}$ (here: $N_{min} = 3$) objects together within one
redshift-slice.

If the EXT-FOF algorithm finds a link between two galaxies, yet the
FOF does not, then one of the following deviations can result:
\begin{enumerate}
  \renewcommand{\theenumi}{(\arabic{enumi})}
\item The group can only be found with EXT-FOF (i.e.\ category 3)
\item The group found with EXT-FOF contains more members than the
  corresponding FOF group (i.e.\ category 5)
\item The EXT-FOF group is a combination of multiple FOF groups
  (i.e.\ category 7)
\end{enumerate}

In the first case, the original FOF might find two linked galaxies,
but cannot find a third object, that fulfills the linking criteria.
Thus the number of objects within the original FOF group-candidate is
too small, i.e.\ eq.\ (\ref{Nminfof}) is violated, and the group is
rejected. If the EXT-FOF method is able to find a third object in the
current redshift-slice, that is linked to one of the galaxies in the
pair, the number of group members satisfies eq.\ (\ref{Nminextfof})
and the group is accepted. It is also possible, that the original FOF
finds two galaxy pairs, but is not capable of connecting them, while
the EXT-FOF algorithm is able to link two galaxies of each pair, thus
creating a group-candidate containing four galaxies. Eq.\ 
(\ref{Nminextfof}) is then satisfied and the EXT-FOF group is
accepted.

In the second case, the EXT-FOF is able to find one or more galaxies,
that are defined as isolated by the FOF algorithm, linked to group
members, resulting in a larger group in the EXT-FOF catalogue.

In the third case, the original FOF finds two or more separate groups.
If the EXT-FOF can find links between members of the different FOF
groups, those groups are combined into one big EXT-FOF structure. Of
course, it is also possible, that further galaxies, defined as
isolated by the original FOF, get attached to such a big structure,
following the reasoning of the second case. This leads to EXT-FOF
groups, that consist of a number of FOF groups and some additional
galaxies.

If the EXT-FOF algorithm cannot find $N_{min}$ galaxies linked
together within at least one of the redshift-slices, one of the
following deviations can result:
\begin{enumerate}
  \renewcommand{\theenumi}{(\arabic{enumi})}
\item The group can only be found with the original FOF (i.e.\ 
  category 2)
\item The group found with FOF contains more members than the
  corresponding EXT-FOF group (i.e.\ category 4)
\item The FOF group is a combination of multiple EXT-FOF groups
  (i.e.\ category 6)
\end{enumerate}

In the first case, the original FOF finds at least $N_{min} = 3$
objects that are linked together, satisfying eq.\ (\ref{Nminfof}). The
EXT-FOF, on the other hand, can only find pairs of galaxies in various
redshift slices. Since every group-candidate has to fulfill eq.\ 
(\ref{Nminextfof}) in at least one redshift slice, these galaxy pairs
are not taken into account in the unification process, and these
galaxies are defined as isolated by the EXT-FOF algorithm.

In the second case, one galaxy, defined as group member by the FOF,
cannot be attached to an EXT-FOF group. This is due to the fact, that
there is no redshift slice, where this galaxy can be linked to at
least two (in general: $N_{min} - 1$) members of this EXT-FOF
structure. Thus the object is not entered into the pre-unification
catalogue and is not included in the EXT-FOF group. It is also
possible, that more than one galaxy of the FOF group is missing in the
corresponding EXT-FOF group. This can happen, if the above mentioned
situation is true for all of those galaxies and, additionally, there
is no redshift slice where two or more of those missing galaxies are
linked with one of the EXT-FOF group members. However, in the case of
the CFA1 group catalogues, none of the five FOF groups, that fall
under category 4, are larger than their EXT-FOF counterpart by more
than one group member.

In the third case, the EXT-FOF finds two or more separate groups,
while the FOF is able to link them together into one structure. This
can happen, if there are no redshift slices, where at least two
members of one EXT-FOF group can be linked with at least one member of
another group. Following the reasoning of the second case, it is
furthermore possible, that the large FOF group can contain additional
members, that are defined as isolated by the EXT-FOF. As far as the
CFA1 catalogues are concerned, this effect never shows up, though.

Category 8 groups are generated by a combination of the two
effects: Members get lost in the EXT-FOF groups, because of too few
linked objects at a given redshift slice, and at the same time new
links can be found to galaxies, that are defined as isolated by the
original FOF, while some subset of the FOF and EXT-FOF group is
identical. 

In one of the cases examined for category 3 of the CFA1 comparison, a
group found with the original FOF, is removed from the group list
because its mean velocity exceeds the limit. The EXT-FOF is able to
find another galaxy linked to this group, pushing the group mean
velocity below the limit. Thus, the group is included in the EXT-FOF,
yet not in the FOF group catalogue. However, this is a border effect
and could in principle also work in the other direction, removing
groups from the EXT-FOF catalogue (i.e.\ category 2).

\begin{figure}
  \includegraphics[width=0.95\columnwidth]{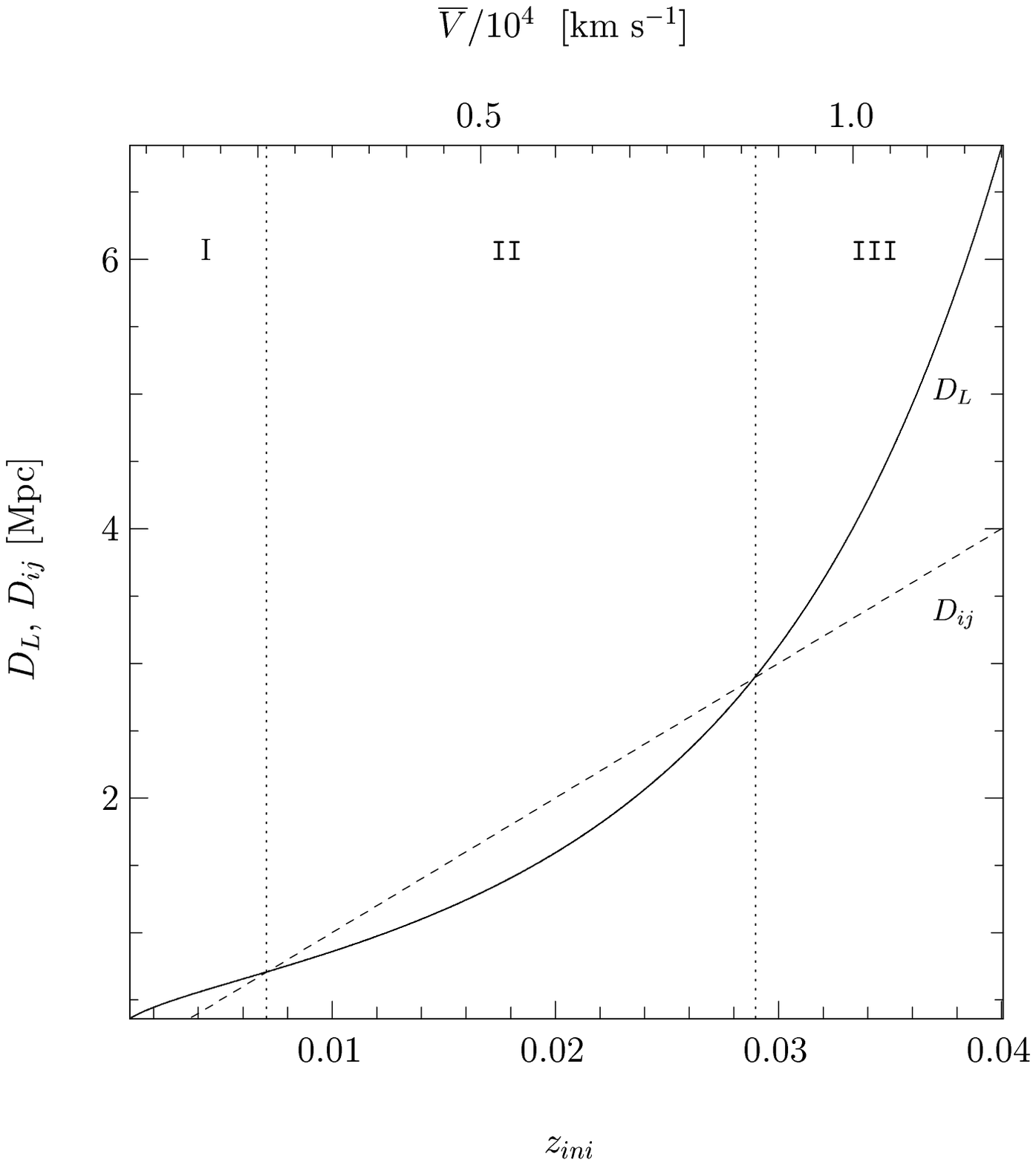}
  \caption{Comparison of the CFA1 projected linking distance $D_L$ (solid
    line) and the projected separation $D_{ij}$ (dashed line) of the
    galaxies $i$ and $j$, as a function of the the mean distance
    $\overline{V}$, or the redshift slice $z_{ini}$, respectively. The
    angular separation $\theta_{ij}$ of the chosen galaxy pair leads
    to two intersections between $D_L$ and $D_{ij}$, dividing the
    graph into three areas (dotted lines). The projected linking
    criterion $D_{ij} \leq D_L$ is satisfied in the areas I and III.
    It is not satisfied in area II.}
  \label{dld12cfa}
\end{figure}

\begin{figure*}
  \begin{minipage}{\textwidth}
    \begin{center}
      \includegraphics[width=7.5cm]{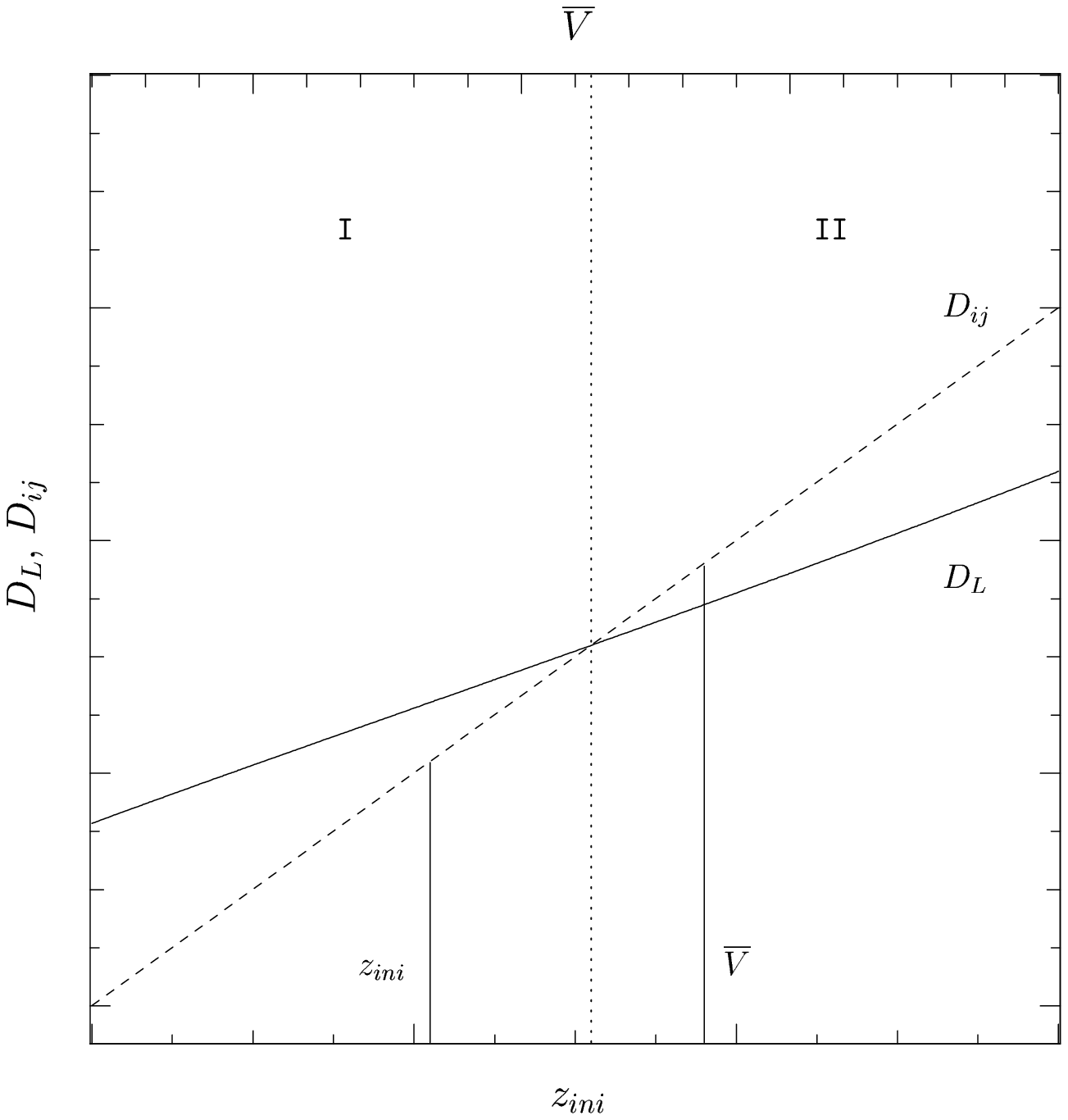} \hspace{1cm}
      \includegraphics[width=7.5cm]{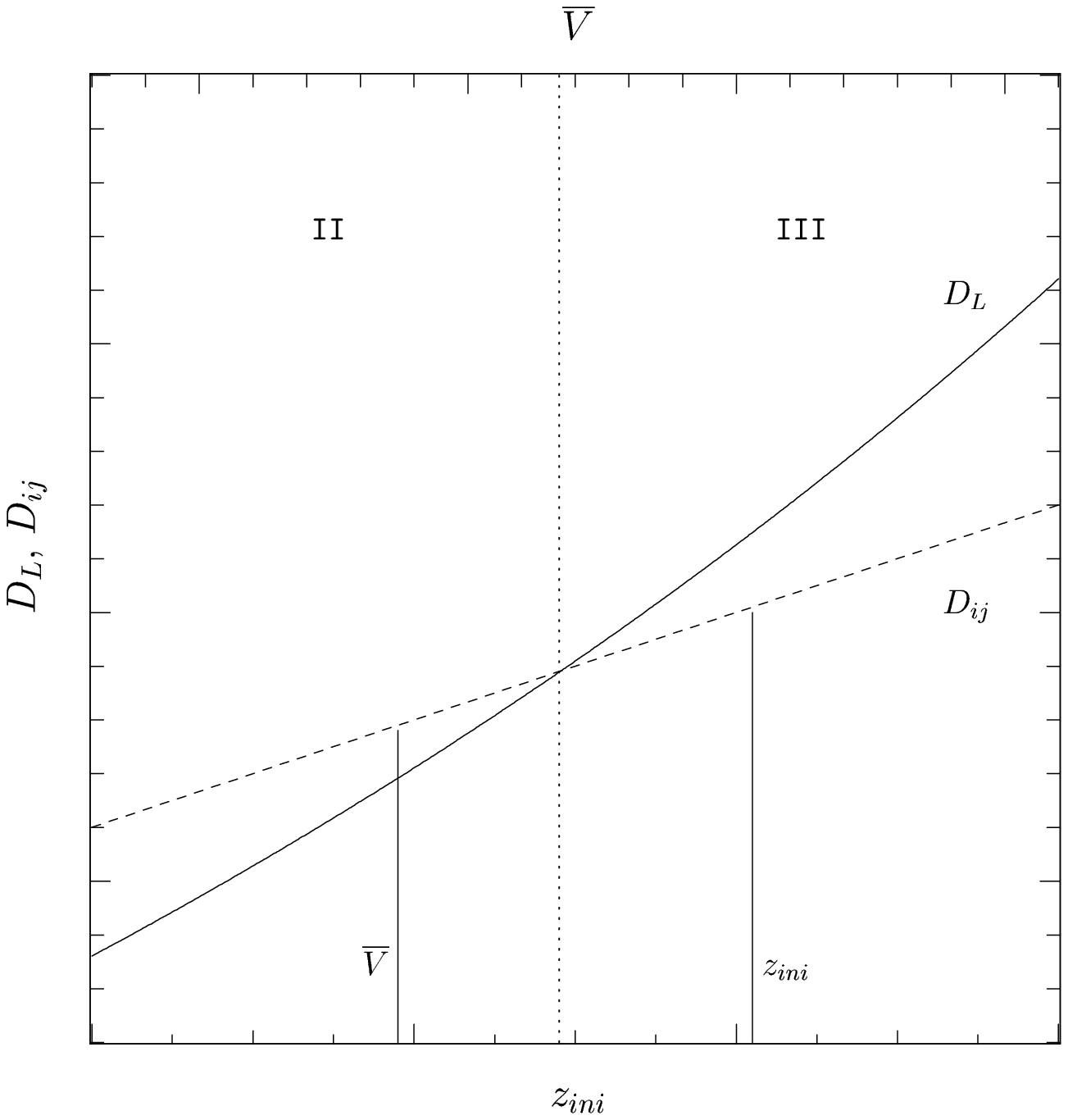}
    \end{center}
    \caption{Comparison of the CFA1 projected linking distance $D_L$ (solid
      line) and projected separation $D_{ij}$ (dashed line), as a
      function of the the mean distance $\overline{V}$, or the
      redshift slice $z_{ini}$, for two example galaxy pairs. These
      schematic drawings show magnified excerpts of the transition
      regions (dotted lines) between the areas I and II (left panel),
      and the areas II and III (right panel), that are described in
      Fig.\ \ref{dld12cfa}. {\bf Left panel:} The projected separation
      $D_{ij}$ at the mean redshift $\overline{V}$ (denoted by the
      solid vertical line in area II) of the galaxies $i$ and $j$,
      does not fulfill the projected linking criterion $D_{ij} \leq
      D_L$. Thus, the original FOF cannot link the two galaxies. In
      the case shown here, the range of $z_{ini}$ values, satisfying
      the EXT-FOF redshift-space linking criteria, eqs.\ 
      (\ref{Vextfofi}) and (\ref{Vextfofj}), reaches into area I. For
      simplification, only one of those redshift slices $z_{ini}$ and
      the corresponding value of $D_{ij}$ is shown here (solid
      vertical line in area I).  $D_{ij} \leq D_L$ is fulfilled for
      this redshift, and the galaxies $i$ and $j$ are found as linked
      by the EXT-FOF algorithm. {\bf Right panel:} $\overline{V}$ is
      again lying in area II, thus the original FOF cannot link this
      galaxy pair. The range of $z_{ini}$ values, that satisfy the
      EXT-FOF redshift-space linking criteria, reach into area III in
      this case. Once more, only one of those redshift slices
      $z_{ini}$ and the corresponding value of $D_{ij}$ is shown here
      (solid vertical line in area III). $D_{ij} \leq D_L$ is
      fulfilled for this redshift, and the galaxy pair $i$ and $j$ are
      found as linked by the EXT-FOF algorithm.}
    \label{dld12abcfa}
  \end{minipage}
\end{figure*}

The reason, why the EXT-FOF algorithm is sometimes able to find links
between objects, while the original FOF is not, can be explained as
follows: 

The equations used here for the calculation of the projected distance
$D_{ij}$, eqs.\ (\ref{Dfof}), (\ref{Dextfof}), and (\ref{Dlowz}), show
that $D_{ij} \propto \overline{V}$ and $D_{ij} \propto z_{ini}$,
respectively. However, being varied with the luminosity function, the
linking distance $D_L$ is not proportional to the redshift. $D_L$ is
growing monotonically with $z$, switching from a convex curvature at
low redshifts, to concave at higher redshifts. The slope of $D_{ij}$
is set by the angular separation $\theta_{ij}$ of the two galaxies.
There are three possible scenarios for the behaviour of $D_{ij}$ and
$D_L$:
\begin{description}
\item Either $\theta_{ij}$ is so small, that $D_{ij} \leq D_L$ over
  the entire redshift range of $\overline{V} \in \left[300\textrm{ km
      s}^{-1}, 12000\textrm{ km s}^{-1}\right]$, or $z_{ini} \in
  \left[0.001, 0.04\right]$ respectively, i.e.\ the projected linking
  criterion, eq.\ (\ref{Dfof}) or (\ref{Dextfof}), is fulfilled at
  $\overline{V}$ and in all other redshift slices $z_{ini}$.
\item Or $\theta_{ij}$ is so large, that $D_{ij} > D_L$ over the
  entire redshift range. This galaxy pair can neither be linked
  together by the FOF, nor by the EXT-FOF algorithm. 
\item Or $\theta_{ij}$ lies between the above mentioned values. In
  this case, $D_{ij}$ intersects $D_L$ twice.
\end{description}
The third scenario can lead to the links, that can only be found with
the EXT-FOF algorithm. In this case, the redshift range can be divided
into three areas, which is shown in Fig.\ \ref{dld12cfa}. The figure
shows a comparison of the projected linking distance $D_L$ (solid
line) and the projected separation $D_{ij}$ (dashed line) of the
galaxies $i$ and $j$, as a function of the the mean distance
$\overline{V}$, or the redshift slice $z_{ini}$, respectively. In the
areas I and III, the projected linking criterion $D_{ij} \leq D_L$ is
fulfilled. Whereas in area II it is never fulfilled. It becomes
obvious, that it is critical in which of those areas the linking
criteria are tested. The original FOF can only test the projected
linking criterion at a fixed redshift $\overline{V}$, set by the
galaxy pair. The EXT-FOF, on the other hand, tests the linking
criteria at all redshifts. If the EXT-FOF redshift-space linking
criteria, eqs.\ (\ref{Vextfofi}) and (\ref{Vextfofj}), can be
fulfilled for a given galaxy pair, then they are generally satisfied
for an interval of $z_{ini}$ values. If eq.\ (\ref{Dextfof}) can also
be satisfied for at least one of those redshifts $z_{ini}$, then all
criteria are fulfilled and the galaxy pair is found as linked by the
EXT-FOF. Thus, only the EXT-FOF and not the original FOF is able to
link a galaxy pair in one of the two following situations: The mean
velocity $\overline{V}$ is lying in the area II, and the redshift
interval of $z_{ini}$ values, that satisfy eqs.\ (\ref{Vextfofi}) and
(\ref{Vextfofj}), reaches either into area I (Fig.\ \ref{dld12abcfa},
left panel), or into area III (Fig.\ \ref{dld12abcfa}, right panel).
As a result, the projected linking criterion is not satisfied for the
original FOF, while both linking criteria can be fulfilled in at least
one $z_{ini}$ slice, in the case of the EXT-FOF.

It can be easily seen, that this effect does not exist, if $D_L$ and
$D_{ij}$ follow the same variation with redshift, $D_L \left(z\right)
\propto D_{ij} \left(z\right)$.

In the following, we explain why the EXT-FOF algorithm sometimes can
not link together $N_{min} = 3$ objects within at least one $z_{ini}$
slice, and why this is the reason for some of the above mentioned
deviations from the original FOF.

The original FOF algorithm can move relatively freely along the
redshift axis, when looking for friends. For example, we consider a
set of galaxies $a$, $b$, and $c$, with given velocities $v_a$, $v_b$,
and $v_c$. For matters of simplicity, we assume $v_a < v_b < v_c$. The
FOF calls those three galaxies a group, if there is at least one
galaxy among them, that can be linked to the other two. If, for
example, the galaxies $a$ and $b$, and $b$ and $c$ fulfill the linking
criteria, i.e.\ eqs.\ (\ref{Dfof}) and (\ref{Vfof}),
then $a$, $b$ and $c$ are linked together. It is not necessary, that
$a$ and $c$ also satisfy the linking criteria, and $V_{ac} > V_L$ is
possible. Thus, in principle, the FOF allows for very elongated chains
of galaxies along the redshift axis, as long as these galaxies have a
small projected separation and the velocity difference $V_{ij}$
between next neighbours satisfies eq.\ (\ref{Vfof}). This effect is
what makes the original FOF unqualified for cluster finding in
datasets with rather uncertain redshifts, like photometric galaxy
catalogues.

The EXT-FOF algorithm cannot move so freely along the redshift axis.
By only looking for friends among galaxies, that are compatible with a
given redshift slice, the probability of finding very elongated
structures is efficiently reduced. 

\begin{figure}
  \includegraphics[width=0.95\columnwidth]{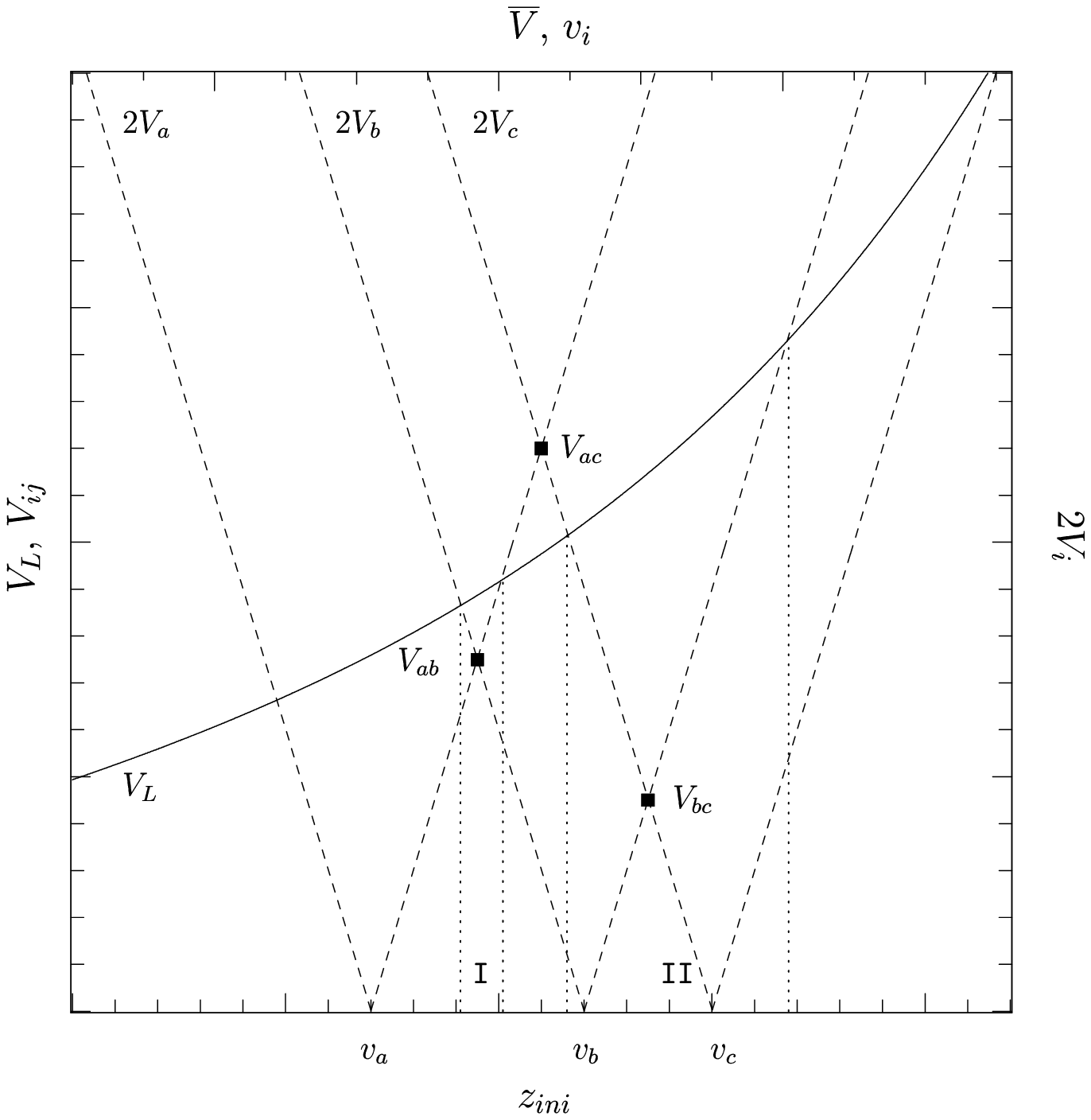}
  \caption{Comparison of the CFA1 linking velocity $V_L$ (solid line) with
    the original FOF (filled squares) and EXT-FOF (dashed lines)
    velocity separations, $V_{ij}$ and $V_i$, for a set of three
    galaxies, $a$, $b$, and $c$, that can only be linked with the
    original FOF, yet not with the EXT-FOF.
    The lower scale shows the redshift $z_{ini}$, the upper scale
    shows the corresponding mean velocity $\overline{V}$. The
    individual velocities $v_a$, $v_b$, and $v_c$ of the three
    galaxies are also shown on the lower scale. The original FOF
    velocity separations $V_{ab}$, $V_{bc}$, and $V_{ac}$, resulting
    from eq.\ (\ref{Vfof}), are plotted at the corresponding mean
    velocities.
    The EXT-FOF velocity separations are plotted as $2\,V_a$,
    $2\,V_b$, and $2\,V_c$ to make an easy comparison with $V_L$
    possible. The dotted lines show the areas, where eqs.\ 
    (\ref{Vextfofi}) and (\ref{Vextfofj}) are satisfied for two
    galaxies: In area I, the requirement is fulfilled for the galaxies
    $a$ and $b$, and in area II it is fulfilled for the galaxies $b$
    and $c$. In no redshift slice, can all $N_{min} = 3$ galaxies
    satisfy the EXT-FOF velocity linking requirement $V_i \leq V_L /
    2$.
    }
  \label{vlcfa}
\end{figure}

To illustrate the problem, Fig.\ \ref{vlcfa} shows the case of a set
of galaxies, $a$, $b$, and $c$, that are found as a group by the FOF,
yet not by the EXT-FOF. For simplification, we choose a galaxy set,
that has a very small spread in the projected distance, so that eqs.\ 
(\ref{Dfof}) and (\ref{Dextfof}) are fulfilled for all three galaxies,
and we only have to test, whether the velocity linking criteria are
fulfilled. The figure shows the variation of the linking velocity
$V_L$ (solid line) with redshift $z_{ini}$, or mean velocity
$\overline{V}$, respectively. The individual velocities $v_a$, $v_b$,
and $v_c$ of the three galaxies are shown on the lower scale. The
filled squares are the original FOF velocity separations $V_{ab}$,
$V_{bc}$, and $V_{ac}$, resulting from eq.\ (\ref{Vfof}), and plotted
at the corresponding mean velocities. It becomes obvious, that
$V_{ab}$ and $V_{bc}$ satisfy the velocity linking criterion, and
together with the already fulfilled projected linking criterion, the
original FOF combines the three galaxies into a group. The dashed
lines denote the EXT-FOF velocity separations. They are plotted as
$2\,V_a$, $2\,V_b$, and $2\,V_c$ to make an easy comparison with $V_L$
possible. The dotted lines show the areas where eqs.\ (\ref{Vextfofi})
and (\ref{Vextfofj}) are satisfied for two galaxies: In area I, the
requirement is fulfilled for the galaxies $a$ and $b$, and in area II
it is fulfilled for the galaxies $b$ and $c$. There is no redshift
slice, where all three galaxies satisfy the EXT-FOF velocity linking
requirement $V_i \leq V_L / 2$. As a result, the EXT-FOF can only find
the galaxies $a$ and $b$, or $b$ and $c$ linked together, in the
$z_{ini}$ ranges given by the areas I, or II. Those galaxy pairs do
not satisfy eq.\ (\ref{Nminextfof}) with $N_{min} = 3 $, and thus are
not added to the group list.

Theoretically, another effect, that could remove a single link between
two galaxies, is conceivable. It would also lead to the observed
phenomena of category 2, 4, 6, or 8. If the step size $\Delta z_{ini}$
of the EXT-FOF redshift slicing is chosen too large, not every
possible value of $\overline{V}$ can be approximated anymore, and a
link found with the original FOF could not be recovered with the
EXT-FOF algorithm. This is never the case in one of the examined
catalogues, though, proving that the chosen step size is sufficiently
small and does not lead to unwanted numerical effects.

\section{Application to the LCRS redshift survey}
\label{s:lcrs}

\subsection{The galaxy catalogue}
\label{ss:lcrssurvey}

The LCRS \citep{Shectman1996} is an $R$-band selected, spectroscopic
galaxy survey. It consists of six $1\fdg5 \times 80\degr$ strips,
three of which are lying in the north, the other three in the south
Galactic cap. The survey covers a total area of over $700\,
\mathrm{deg}^2$, divided into 327 individual fields. For 120 of those
fields spectroscopy is done with a 50 fibre multiobject spectrograph
(MOS), and they have nominal apparent magnitude limits of
\begin{equation}
16.0\;\textrm{mag} \leq m_R \leq 17.3\;\textrm{mag} .
\end{equation}
For the remaining 207 fields spectroscopy is done with a 112 fibre
MOS, with nominal apparent magnitude limits of
\begin{equation}
15.0\;\textrm{mag} \leq m_R \leq 17.7\;\textrm{mag} .
\end{equation}
For every setup, all of the fibres are used, but since each of the
fields is observed only once, there is a field-to-field variation in
the selection criteria. The protective tubing around the individual
fibres of the MOS makes it impossible to observe spectra of objects,
that are closer to each other than $55\arcsec$, leading to $55\arcsec$
``orphans'', that have no spectroscopic redshifts. The surveyed
galaxies are sampled randomly within each field.

The catalogue we use here is a combination of the 23695 galaxies
that have spectroscopic redshifts and the 1694 $55\arcsec$
``orphans'', resulting in a total of 25389 galaxies. Both datasets
only include objects, that are lying within the geometric and
photometric boundaries of the survey. Artificial redshifts are
assigned to the $55\arcsec$ ``orphans'', by giving each of them the
redshift of its nearest neighbour, convolved with a Gaussian of width
$\sigma = 200 \textrm{ km s}^{-1}$ (\citealt{Tucker2000}, Tucker,
priv.\ comm.).

\subsection{Creation of the group catalogues}
\label{ss:lcrsgroups}

Following the recipe of \citet{Tucker2000}, we once more create two
structure catalogues, one with the original FOF algorithm, the
other with our EXT-FOF. In the following, an overview of the
\citet{Tucker2000} treatment of the data is shown. Furthermore, the
application of this recipe to the EXT-FOF is explained.

First, all galaxy velocities are corrected for motion relative to the
dipole moment of the cosmic microwave background. The equation used
by \citet{Tucker2000} is given in galactic coordinates
\begin{equation}
V_V = V_\odot \left[\sin b_i \sin b_\odot + \cos b_i 
\cos b_\odot \cos\left(l_i - l_\odot \right)\right] ,
\end{equation}
where $V_\odot = 368.9 \textrm{ km s}^{-1}$, $b_\odot = 48\fdg05$, and
$l_\odot = 264\fdg33$ (\citealt{Lineweaver1996}, Tucker, priv.\ comm.),
and $b_i$, $l_i$ are the galactic latitude and longitude of galaxy
$i$.

The set of galaxies used for cluster finding is limited to objects
having corrected velocities $v_i$ such that 
\begin{equation}
7500 \textrm{ km s}^{-1} \leq v_i < 50000 \textrm{ km s}^{-1} ,
\end{equation}
and absolute magnitudes $M_i$ between
\begin{equation}
-22.5\;\textrm{mag} + 5 \log h \leq M_i < -17.5\;\textrm{mag} + 5 \log
    h .
\end{equation}
Not taking any colour corrections into account, we simply set
\begin{equation}
\label{Mimi}
M_i = m_i - 5 \log \left(d_L \left(H_0, \Omega_M, \Omega_\Lambda,
    v_i/c \right)\right) + 5 ,
\end{equation}
where $m_i$ is the apparent magnitude of galaxy $i$, and $d_L
\left(H_0, \Omega_M, \Omega_\Lambda, v_i/c \right)$ is the luminosity
distance for the given cosmology to galaxy $i$ \citep{CPT:1992}. Since
the LCRS is relatively deep, going out to redshifts of roughly 0.2, it
becomes necessary to use cosmologically correct expressions.

Only groups with mean velocities $\overline{V_g}$ such that
\begin{equation}
10000 \textrm{ km s}^{-1} \leq \overline{V_g} < 45000 \textrm{ km
  s}^{-1} 
\end{equation}
are accepted in the final group catalogue.

We use a flat cosmology with $H_0 = 65 \textrm{ km s}^{-1}\textrm{
  Mpc}^{-1}$, $\Omega_M = 0.3$, and $\Omega_\Lambda = 0.7$.


The distance equations for the original FOF have to be modified to
take the cosmologically correct expressions into account. Eq.\
(\ref{Dfof}) becomes
\begin{equation}
\label{Dfoflcrs}
D_{ij} = 2 \sin \frac{\theta_{ij}}{2} D_{ave} 
\leq D_L ,
\end{equation}
with the mean comoving angular distance $D_{ave}$ between the galaxies
$i$ and $j$
\begin{equation}
\label{Davelcrs}
D_{ave} := \frac{d_A \left(H_0, \Omega_M, \Omega_\Lambda, \frac{v_i}{c}
  \right) + d_A \left(H_0, \Omega_M, \Omega_\Lambda, \frac{v_j}{c}
  \right)}{2} 
\end{equation}
replacing the $\overline{V}$ of eq.\ (\ref{overV}).

In the case of the EXT-FOF, we use $D\left(z_{ini}\right)$ of eq.\ 
(\ref{Dhighz}) for the calculation of the projected distance $D_{ij}$.
The individual spectroscopic redshift errors of the galaxies are not
yet taken into account, thus eqs.\ (\ref{Vextfofi}) and
(\ref{Vextfofj}) still apply. A stepsize of $\Delta z_{ini} = 10^{-5}$
is used.

The linking parameters are varied with redshift. Yet, due to the
field-to-field variations in the sampling fraction, the photometric
limits, and the number of MOS fibres, this dependency is more
complicated than in the case of the CFA1. Eqs.\ (\ref{varDL}) and
(\ref{varVL}) are used for the variation of $D_L$ and $V_L$, and $R$
is given by
\begin{equation}
R =
\left[\frac{n^{exp}\left(f,D\right)}
{n^{exp}_{fid}}\right]^{-1/3} .
\end{equation}
$n^{exp}\left(f,D\right)$ is the expected galaxy number density in the
field $f$ at a comoving distance $D$. $D$ is either $D_{ave}$ in the
case of the original FOF, or $D\left(z_{ini}\right)$ in the case of
the EXT-FOF. $n^{exp}_{fid}$ is the galaxy number density
$n^{exp}\left(f,D\right)$ in a fiducial field at a given fiducial
distance. $n^{exp}\left(f,D\right)$ is calculated with the help of the
Schechter luminosity function $\Phi\left(M\right)$:
\begin{equation}
n^{exp}\left(f,D\right) = F \int_{M_{min}}^{M_{max}}
\Phi\left(M\right) dM ,
\end{equation}
where $F$ is the sampling fraction of field $f$. $M_{min}$ and
$M_{max}$ are the absolute photometric limits of this field, at the
given comoving distance $D$. When comparing two galaxies from
two different fields $f_a$, and $f_b$, the average of the expected
galaxy densities is taken:
\begin{equation}
n^{exp}\left(f,D\right) =
\frac{n^{exp}\left(f_a,D\right)+n^{exp}\left(f_b,D\right)}{2} .
\end{equation}
Two Schechter luminosity functions are used \citep{LCRS96a,
  Tucker1997}, following the approach of \citet{Tucker1997}. One is
valid for the 50 fibre fields in the southern Galactic cap, and has
$\alpha = -0.74$, $M^* = -20.55 \, \textrm{mag} + 5 \log h$, and
$\Phi^* = 0.016 \,h^3\textrm{Mpc}^{-3}$. The other is used for the 50
fibre fields in the northern Galactic cap, as well as for all 112
fibre fields. Its parameters are $\alpha = -0.70$, $M^* = -20.29 \,
\textrm{mag} + 5 \log h$, and $\Phi^* = 0.019 \,h^3\textrm{Mpc}^{-3}$.

For the fiducial field the latter luminosity function is used.
Furthermore, this field has a sampling fraction of $F=1$, and apparent
magnitude limits of
\begin{equation}
15.0\;\textrm{mag} \leq m_R \leq 17.7\;\textrm{mag} .
\end{equation}
The fiducial redshift is set to $c\,z_{fid} = 30000 \textrm{ km
  s}^{-1}$.

Only groups having at least $N_{min} = 3$ members are searched for.
The projected separation $D_0$ and the velocity difference $V_0$ are
chosen as $D_0 = 0.715 h^{-1}\textrm{Mpc}$ and $V_0 = 500 \textrm{ km
  s}^{-1}$, respectively, again following \citet{Tucker1997}.

\begin{figure}
  \includegraphics[width=\columnwidth]{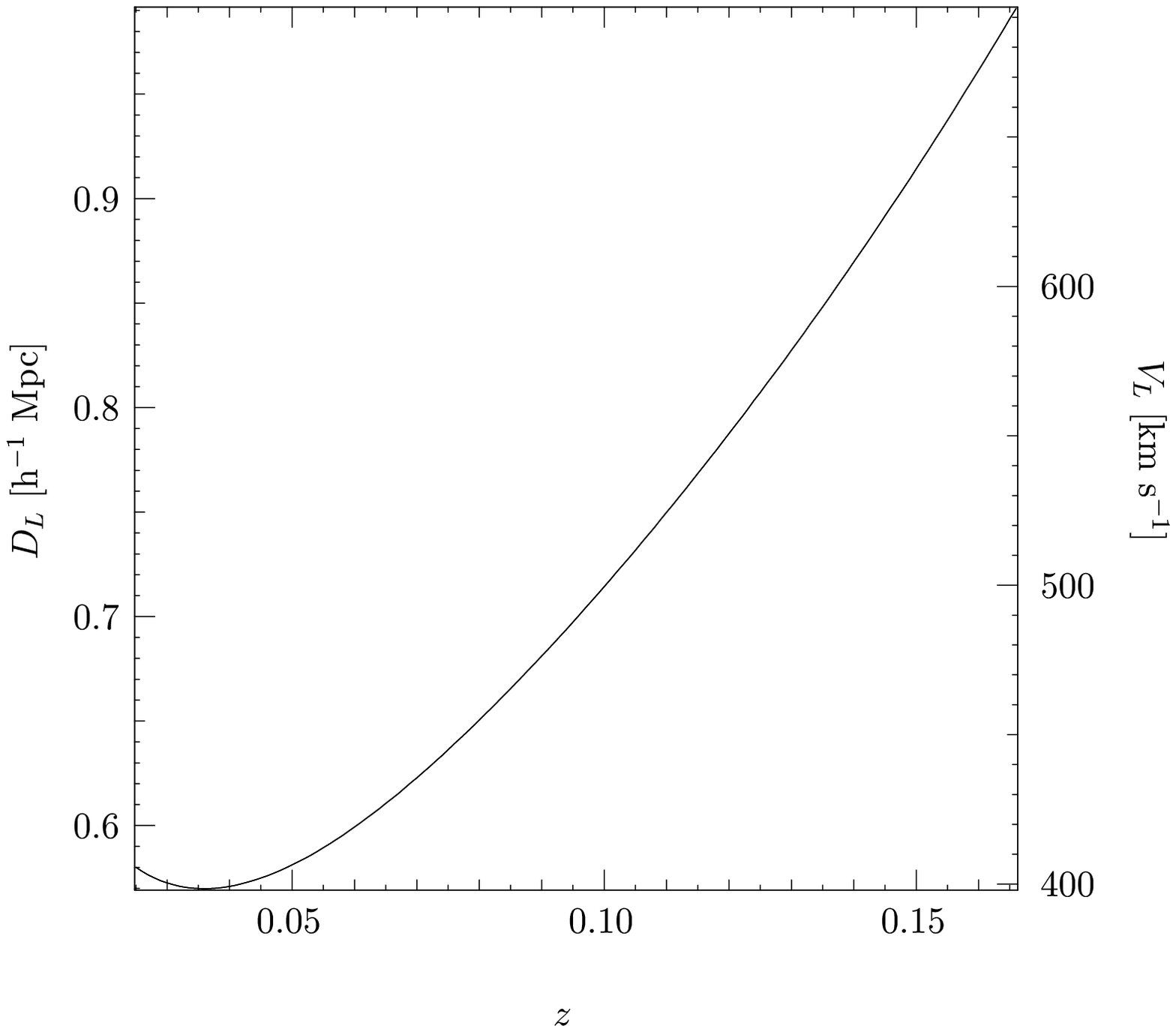}
  \caption{Variation of the linking criteria $D_L$ and $V_L$ for the
    LCRS group catalogues, as a function of
    redshift $z$. 
    As a reminder, in this case $R$ is not only a function of
    distance, but also depends on the field. We show a typical example
    field, having a MOS setup with 112 fibres, a sampling fraction of
    $F = 0.7$, and photometric limits of $m_{min} =
    15.0\;\textrm{mag}$, $m_{max} = 17.7\;\textrm{mag}$.}
  \label{dlvllcrs}
\end{figure}

The resulting variation of the linking criteria $D_L$ and $V_L$ as a
function of
redshift $z$ is shown in Fig.\ \ref{dlvllcrs}. This graphic is created
for an example field having a MOS setup with 112 fibres and
photometric limits $m_{min} = 15.0\;\textrm{mag}$, $m_{max} =
17.7\;\textrm{mag}$. We use a typical sampling fraction of $F = 0.7$.

\subsection{Comparison of the group catalogues}
\label{ss:lcrscomp}

\begin{table*}
  \caption{Comparison of the LCRS FOF and EXT-FOF structures}
  \label{tablelcrs}
  \begin{tabular}{lccccc}
    \hline
    & category & FOF & EXT-FOF & \% FOF & \% EXT-FOF \\
    \hline
    total &  & 1367 & 1285 & 100 & 100 \\
    &&&&& \\
    identical & 1 & 1145 & 1145 & 83.8 & 89.1 \\
    only FOF struct.\ & 2 & 87 & -- & 6.4 & -- \\
    only EXT-FOF struct.\ & 3 & -- & 1 & -- & 0.1 \\
    FOF struct.\ larger than EXT-FOF & 4 & 120 & 120 & 8.8 & 9.3 \\
    EXT-FOF struct.\ larger than FOF & 5 & 4 & 4 & 0.3 & 0.3 \\
    FOF struct.\ is combination of EXT-FOF structs.\ & 6 & 5 & 10 & 0.4 &
    0.8 \\
    EXT-FOF struct.\ is combination of FOF structs.\ & 7 & 2 & 1 & 0.2
    & 0.1 \\
    FOF and EXT-FOF structs.\ have some elements in common & 8 & 4 & 4 & 0.3 & 0.3 \\
    &&&&& \\
    FOF structs.\ found with EXT-FOF alg.\ &  & 1280 & -- & 93.6 & -- \\
    \hline
  \end{tabular}

\end{table*}

Application of the original FOF yields a total of 1367 groups. 6747
galaxies are contained in these groups. The structures are slightly
deviant from the ones published by \citet{Tucker2000}. These
discrepancies in the group composition arise from three differences in
the treatment of the data: First, we include no colour corrections in
eq.\ (\ref{Mimi}), unlike \citet{Tucker2000}. Second, for the
determination of $D_{ave}$ and $D\left(z_{ini}\right)$ we make use of
the comoving angular distance $d_A \left(H_0, \Omega_M,
  \Omega_\Lambda, z \right)$, while \citet{Tucker2000} utilised the
proper motion distance $d_M \left(H_0, \Omega_M,
  \Omega_\Lambda, z \right)$. Third, we determine group memberships in
a flat universe with a low matter content, whereas \citet{Tucker2000}
used an Einstein-de Sitter universe.

The catalogue resulting from the EXT-FOF application contains 1285
groups, with a total of 6337 galaxies. 

Table \ref{tablelcrs} shows the statistics of the comparison between
the FOF and EXT-FOF structures in the case of the LCRS. The same
object-to-object comparison is used as described in Section
(\ref{ss:cfacomp}).

1280 of the FOF groups are recovered with the EXT-FOF algorithm,
corresponding to a recovery rate of almost 94\%. The EXT-FOF, on the
other hand, finds only one group, that is not a member of the FOF
group catalogue (category 3), leading to a spurious detection rate of
less than 1\%. 
\cbstart
Both the recovery and spurious detection rates are based on the
assumption that the original FOF algorithm used here yields a complete
structure catalogue.
\cbend
1145 structures are found identical by both algorithms,
so the rate of identical recoveries lies at roughly 86\%.  Of the 87
FOF groups, that have no EXT-FOF counterparts (category 2), 77 have
only three members, and the remaining ten contain four objects each.
Those groups are all very elongated in redshift, making their group
status questionable, anyway. The one group falling under category 3 is
also extremely small, having only three members.  The above comparison
shows, that both algorithms yield very similar results.

\subsection{Analysis of the discrepancies}
\label{ss:lcrsanalysis}

Like in the case of the CFA1, all of the deviations mentioned above
can be explained by the two effects, that are described in Section
\ref{ss:cfaanalysis}.

\begin{figure}
  \includegraphics[width=0.95\columnwidth]{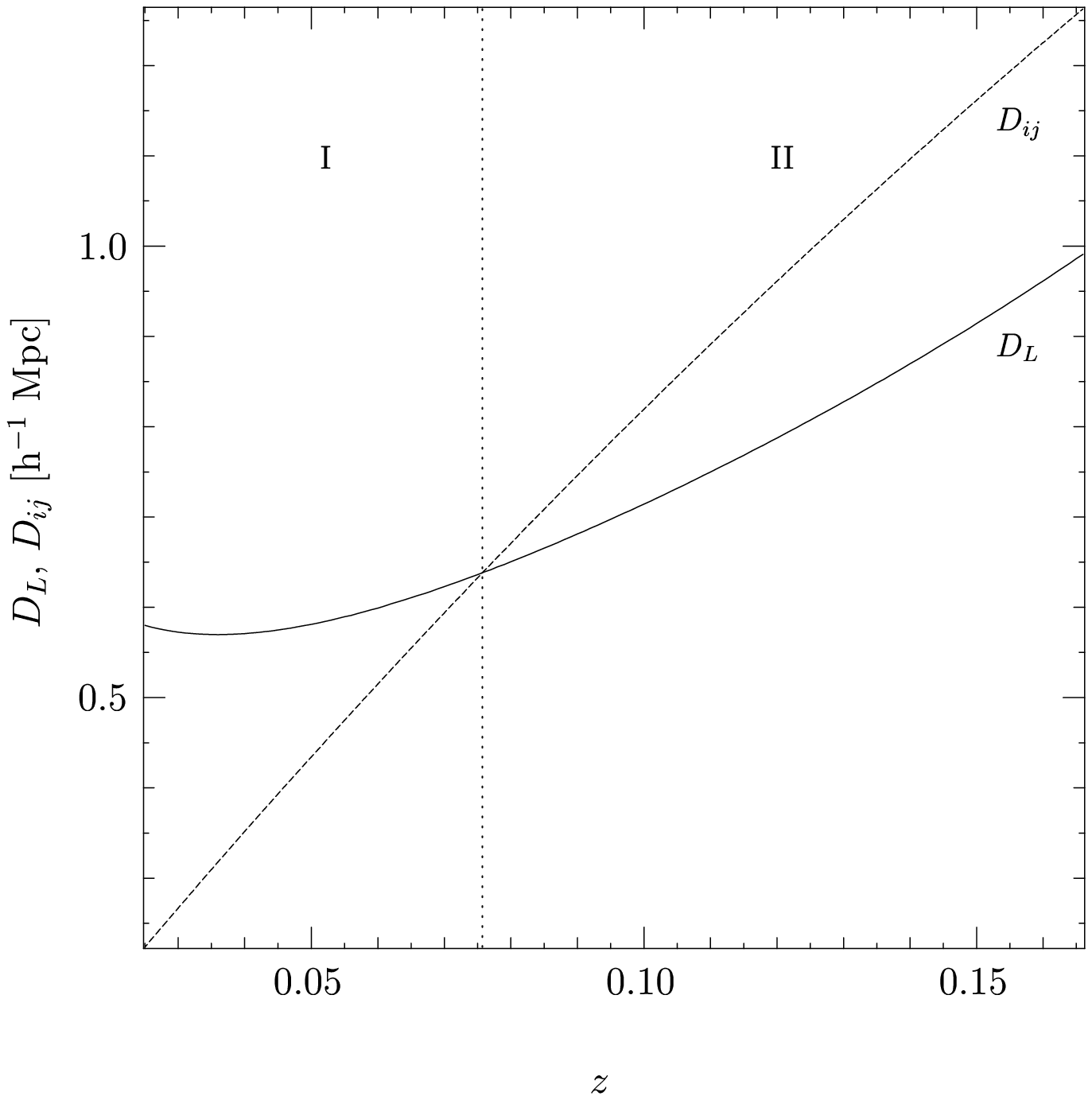}
  \caption{Comparison of the LCRS projected linking distance $D_L$ (solid
    line) and the projected separation $D_{ij}$ (dashed line) of the
    galaxies $i$ and $j$, as a function of the redshift $z$. The
    angular separation $\theta_{ij}$ of the chosen galaxy pair leads
    to one intersection between $D_L$ and $D_{ij}$, dividing the graph
    into two areas (dotted line). The projected linking criterion
    $D_{ij} \leq D_L$ is satisfied in area I, yet not in area II. This
    example galaxy pair is lying in a field, with a MOS setup of 112
    fibres, a sampling fraction of $F = 0.7$, and photometric limits
    of $m_{min} = 15.0\;\textrm{mag}$, $m_{max} = 17.7\;\textrm{mag}$.}
  \label{dld12lcrs}
\end{figure}

The reason for finding an additional link with the EXT-FOF in the LCRS
catalogue is very similar to the case of the CFA1. The equations used
for the determination of the projected distance $D_{ij}$, eqs.\ 
(\ref{Dfoflcrs}), (\ref{Davelcrs}), and
(\ref{Dextfof}), (\ref{Dhighz}), show that $D_{ij}$ grows with $d_A
\left(H_0, \Omega_M, \Omega_\Lambda, z \right)$, while the linking
distance $D_L$ is varied with the luminosity function. $D_{ij}$ has a
convex curvature, $D_L$ is concave. The slope of $D_{ij}$ is once more
set by the angular separation $\theta_{ij}$, and there are again three
possible scenarios for the behaviour of $D_{ij}$ and $D_L$. Very small
angular separations lead to $D_{ij} \leq D_L$, and very large values
of $\theta_{ij}$ lead to $D_{ij} > D_L$ for the entire redshift range,
that is used for cluster finding. Fig.\ \ref{dld12lcrs} shows the case
of an example galaxy pair with a medium sized angular separation,
leading to one intersection between $D_{ij}$ (dashed line) and $D_L$
(solid line). The intersection divides the graph into two areas
(dotted line). In area I the projected linking criterion $D_{ij} \leq
D_L$ is satisfied, in area II it is not. Thus, whenever the original
FOF is limited to comparing $D_{ij}$ with $D_L$ at a value of
$D_{ave}$, that corresponds to a redshift in area II, and the
redshift-space linking criteria, eqs.\ (\ref{Vextfofi}) and
(\ref{Vextfofj}), allow the EXT-FOF to compare $D_{ij}$ with $D_L$ in
area I, only the EXT-FOF can find a link between the galaxies $i$ and
$j$.

The explanation, why the EXT-FOF cannot link together certain objects
in the LCRS catalogue, while the original FOF can, is completely
analogous to the case of the CFA1: The original FOF can move
relatively freely through redshift space in order to link galaxies to
one another. The EXT-FOF, on the other hand, is always limited to
fulfilling both linking criteria for at least $N_{min}$ objects within
one redshift slice.

\section{LCRS with simulated photometric redshifts}
\label{s:lcrsphotred}

In order to prove the validity of the EXT-FOF algorithm for structure
finding in photometric redshift datasets, we create artificial
photometric redshifts for all galaxies contained in our LCRS galaxy
catalogue. We determine a group catalogue, based on this new dataset,
with the help of the EXT-FOF algorithm, taking into account the
photometric redshifts and their errors. A comparison between this
structure catalogue and the EXT-FOF catalogue described in Section
\ref{ss:lcrsgroups} shows the applicability of the EXT-FOF technique
in case of galaxy catalogues with photometric redshifts.

\subsection{Creation of the galaxy dataset with simulated photometric 
redshifts}
\label{ss:lcrssimzgals}

For every LCRS galaxy a random redshift offset is created. The offsets
follow a Gaussian distribution with a $\sigma$ of roughly 5\% of the
survey depth. That percentage resembles the typical proportions
between the redshift errors and the depth of a photometric redshift
survey. In the case of the LCRS, the width of the Gaussian is set to
$c\sigma = 2500 \textrm{ km s}^{-1}$. The photometric redshifts are
then created by adding the individual offsets to the spectroscopic
redshifts of the galaxies. Furthermore, the redshift errors $\delta
z_i$ are all set to $\sigma$. Except for the redshifts of the galaxies
and their errors, the galaxy dataset remains unchanged.

\subsection{Creation of the group catalogue}
\label{ss:lcrssimzgroups}

To construct the EXT-FOF structure catalogue of this
pseudo-photometric redshift dataset, we follow the recipe described in
Section \ref{ss:lcrsgroups}.

The new galaxy dataset is given the same treatment as the
spectroscopic one, as far as the velocity correction, and the culling
of galaxies on the basis of the velocity and absolute magnitude limits
is concerned. Due to this selection process, the subset of galaxies
that enters the structure finding procedure can be deviant from the
one used in the spectroscopic redshift case. A total of 22739 galaxies
are used for cluster finding in the spectroscopic redshift case, 22617
galaxies are used in the photometric redshift case. 20255 galaxies are
common to both culled datasets.

For the calculation of the projected distance $D_{ij}$, we use
$D\left(z_{ini}\right)$ of eq.\ (\ref{Dhighz}). The relatively large
redshift errors are taken into account, and relations
(\ref{Vextfoferri}) and (\ref{Vextfoferrj}) apply for the linking
velocities $V_{L,i}$ and $V_{L,j}$. The stepsize $\Delta z_{ini}$ is
again set to $10^{-5}$.

The same low matter density cosmology is used as described in Section
\ref{ss:lcrsgroups}. We furthermore use the same linking parameters
and group mean velocity limits. The minimum number of objects is once
more set to $N_{min} = 3$.

\subsection{Comparison of the group catalogues}
\label{ss:lcrssimzcomp}

\begin{table*}
  \caption{Comparison of the LCRS EXT-FOF structure catalogues with and
    without simulated photometric redshifts}
  \label{tablelcrssim}
  \begin{tabular}{lccccc}
    \hline
    & category & EXT-FOF & EXT-FOF & \% EXT-FOF & \% EXT-FOF \\
    & & without $\delta z$ & with $\delta z$ & without $\delta z$ & with $\delta z$ \\
    \hline
    total &  & 1285 & 1598 & 100 & 100 \\
    &&&&& \\
    identical & 1 & 230 & 230 & 17.9 & 14.4 \\
    only EXT-FOF without $\delta z$ struct.\ & 2 & 94 & -- & 7.3 & -- \\
    only EXT-FOF with $\delta z$ struct.\ & 3 & -- & 646 & -- & 40.4 \\
    EXT-FOF without $\delta z$ struct.\ larger than EXT-FOF with
    $\delta z$ & 4 & 34 & 34 & 2.7 & 2.1 \\
    EXT-FOF with $\delta z$ struct.\ larger than EXT-FOF without
    $\delta z$ & 5 & 382 & 382 & 29.7 & 23.9 \\
    EXT-FOF without $\delta z$ struct.\ is combination of EXT-FOF with
    $\delta z$ structs.\ & 6 & 0 & 0 & 0.0 & 0.0 \\
    EXT-FOF with $\delta z$ struct.\ is combination of EXT-FOF without
    $\delta z$ structs.\ & 7 &  229 & 92 & 17.8 & 5.8 \\
    EXT-FOF with and without $\delta z$ structs.\ have some elements
    in common & 8 & 316 & 214 & 24.6 & 13.4  \\
    &&&&& \\
    EXT-FOF without $\delta z$ structs.\ found with EXT-FOF with
    $\delta z$ alg.\ &  & 1191 & -- & 92.7 & -- \\
    \hline
  \end{tabular}

\end{table*}

Section \ref{ss:lcrscomp} shows that in case of the spectroscopic
dataset the EXT-FOF algorithm yields a group catalogue, that is very
similar to the FOF group catalogue. Thus, in principle, the groups
resulting from the photometric redshift dataset could be compared to
either the FOF, or the EXT-FOF group catalogue of the spectroscopic
dataset. For the following examination we choose the latter.

The EXT-FOF algorithm finds a total of 1598 structures in the
simulated photometric redshift galaxy catalogue. The groups and
clusters contain 10831 galaxies. This corresponds to a mean of almost
seven objects per structure. In the spectroscopic redshift case,
structures contain a mean of almost five objects. Obviously, the
EXT-FOF algorithm tends to find more and larger structures in
photometric redshift datasets.

The comparison between the EXT-FOF structures resulting from the
spectroscopic redshift dataset and the structures resulting from the
photometric one is shown in table \ref{tablelcrssim}. Again, we use
the ``category 1 - 8'' classification scheme, that we already
described in Sect.\ \ref{ss:cfacomp}, replacing ``FOF'' with ``EXT-FOF
with spectroscopic redshift dataset'' and ``EXT-FOF'' with ``EXT-FOF
with photometric redshift dataset''.

Only 94 of the 1285 EXT-FOF structures with spectroscopic redshifts
cannot be retrieved (category 2), leading to a recovery rate of
almost 93\%. 646 additional groups are found in the photometric galaxy
dataset (category 3), corresponding to a spurious detection rate of
40\%. 
\cbstart
The recovery and spurious detection rates are based on the
assumption that the application of the EXT-FOF algorithm to the
spectroscopic dataset results in a complete structure catalogue. The
similarity between the original FOF groups and the EXT-FOF groups in
the case of the spectroscopic galaxy dataset was already confirmed in
Sect.\ \ref{ss:lcrscomp}.
\cbend
The number of structures, that are identical in both group
catalogues is roughly 16\%. Due to the strongly differing input galaxy
catalogues, this small percentage does not come as a surprise.  Both
the category 2 and 3 structures tend to be relatively small. 96.8\% of
the category 2 and 81.6\% of the category 3 groups have less than five
members.

Under the assumption that the structures found by the FOF and
EXT-FOF algorithms in the spectroscopic galaxy dataset are real, this
comparison shows that the EXT-FOF algorithm is capable of finding
almost all of the structures contained in this photometric redshift
dataset. Roughly 60\% of the found structures can be expected to be
real. This demonstrates that the EXT-FOF technique is a very
conservative cluster finder and can be used for identification of
cluster candidates in photometric redshift galaxy datasets.

\subsection{Analysis of the discrepancies}
\label{ss:lcrssimzanalysis}

Some of the discrepancies in the two group catalogues are due to the
slightly different composition of the culled input galaxy datasets.
After all, the composition of those datasets differs by roughly 10\%.
However, since this is only a border effect, we do not wish to give it
further attention here.

The really interesting discrepancies can be ascribed to either one, or
both, of the following two reasons:
\begin{description}
\item One is the scattering of the newly created photometric redshifts
  against the real, spectroscopic ones.
\item The other is the fact that the redshift errors are finally
  taken into account when testing the velocity linking criteria,
  using relations (\ref{Vextfoferri}) and (\ref{Vextfoferrj}).
\end{description}

Due to the Gaussian nature of the simulated photometric redshift
distribution, almost 32\% of the galaxies have new redshifts, that
deviate by more than $1 \sigma$ from their original spectroscopic
ones. However, eqs.\ (\ref{Vextfoferri}) and (\ref{Vextfoferrj}) only
allow for a deviation from $z_{ini}$ of roughly $1 \sigma$ ($V_L / 2$
is relatively small compared to $c \delta z_i = c \delta z_j = c
\sigma$). So there is a non-negligible amount of galaxies in the
photometric redshift dataset, that are found as linked in the
spectroscopic redshift cluster catalogue, yet cannot be linked
together anymore, because of their large scatter in photometric
redshift. On the other hand, this scatter can also link galaxies
together, that are defined as separated in the spectroscopic redshift
dataset. So the scattering is responsible for both, finding and losing
links between galaxies, and can result in any type of deviation
ranging from category 2 to 8.
 
Taking the redshift errors into account leads to much larger
velocity linking criteria $V_{L,i}$ and $V_{L,j}$, than in the
spectroscopic redshift case where we use $V_L/2$ as linking
criterion. Thus, the new velocity linking criteria are by far less
strict, and the probability of finding two objects as linked is
increased. The new velocity linking criteria are responsible for
finding additional links and result in deviations of category 3, 5,
and 7. In combination with the above mentioned scattering of
photometric redshifts they can also result in category 8
discrepancies. The consideration of the photometric redshift errors is
the main reason for the obviously larger and more numerous structures
found in the photometric redshift structure catalogue.

\section{Summary}
\label{s:summary}

In this paper we presented a new structure finding algorithm designed
to detect groups and clusters of galaxies in a galaxy catalogue having
photometric redshifts. 

Since the development of the photometric redshift technique this type
of galaxy survey has grown in popularity. The relatively large errors
in the redshift evaluation pose a problem for a reliable cluster
identification, though. One of the most common cluster finding methods
is the friends-of-friends (or FOF) technique. It utilises a
straightforward approach by looking for overdensities in the galaxy
distribution and not depending on any model assumptions for clusters,
and it proved to yield reliable results. However this algorithm is
only capable of looking for structures in either spectroscopic
redshift datasets, or simulated galaxy datasets. We thus created a
modified algorithm, the extended friends-of-friends (or EXT-FOF),
based on the original FOF, that is a tailor-made solution for dealing
with the large photometric redshift errors.

We presented our new algorithm and explained the major differences
between the original FOF and the EXT-FOF method. The basic working
principle of the EXT-FOF being the limitation to galaxies for
structure finding, that are compatible within their redshift errors
with given redshift values.

The EXT-FOF algorithm was tested on two spectroscopic redshift
surveys, the Center for Astrophysics Redshift Survey (CFA1) and the
Las Campanas Redshift Survey (LCRS). The resulting groups were
compared to structures found with the original FOF in those two
surveys. We were able to show that in case of spectroscopic redshift
datasets, the FOF and EXT-FOF group catalogues were almost identical
to each other. The comparison provided a recovery rate of almost 98\%
and 94\% of the FOF structures for the CFA1 and LCRS cluster
catalogues, respectively. A spurious detection rate of less than 5\%
and 1\% of the EXT-FOF structures was found for the CFA1 and LCRS
cluster catalogues. 
\cbstart
Both the recovery and spurious detection rates were based on the
assumption that the original FOF algorithm yielded a complete
structure catalogue. 
\cbend
All of the discrepancies between the FOF and
EXT-FOF group compositions were explained. We showed that our new
algorithm is reluctant to identify an accumulation of galaxies as a
cluster if it is very elongated along the redshift axis. This is an
attribute, that makes the EXT-FOF technique very valuable for
cluster-finding in photometric redshift datasets.

We furthermore tested our EXT-FOF algorithm on a dataset with
simulated photometric redshifts that we created by assigning Gaussian
distributed redshift offsets to the galaxies of the LCRS. A comparison
of the resulting group catalogue with the one found by the EXT-FOF in
case of the spectroscopic LCRS proved, that the EXT-FOF method yields
a conservative cluster catalogue, and is very capable of structure
finding in photometric redshift galaxy datasets. A recovery rate of
almost 93\% of the spectroscopic redshift structures and a spurious
detection rate of about 40\% of the photometric redshift structures
were found. Furthermore, as was expected, a general tendency for
finding larger groups was discovered in the photometric redshift case.

This paper is the first in a series of papers, dealing with the search
for groups and clusters of galaxies in photometric redshift datasets.
The next paper will show the application of the new EXT-FOF algorithm
to the Munich Near-IR Cluster 
Survey (MUNICS; Drory et al.\ 2001b).

\section*{Acknowledgements}

We want to thank Dr.\ John Huchra cordially for discussing his galaxy
catalogue with us. Further sincere thanks go to Dr.\ Douglas Tucker
for his extensive comments by e-mail. The MUNICS project was supported
by the Deutsche Forschungsgemeinschaft,
\textit{Sonderforschungsbereich 375, Astroteilchenphysik}.

\nocite{Drory2001}
\nocite{MUNICS1}
\nocite{McC012}
\nocite{McC011}
\nocite{combo17}
\nocite{CADIS1}

\bibliography{mnrasmnemonic,literature} \bibliographystyle{mn2e}

\begin{thebibliography}{}

\bibitem[\protect\citeauthoryear{{Abell}}{{Abell}}{1958}]{Abell1958}
{Abell} G.~O.,  1958, ApJS, 3, 211

\bibitem[\protect\citeauthoryear{{Abell}, {Corwin} \& {Olowin}}{{Abell}
  et~al.}{1989}]{ACO1989}
{Abell} G.~O.,  {Corwin} H.~G.,    {Olowin} R.~P.,  1989, ApJS, 70, 1

\bibitem[\protect\citeauthoryear{{Bahcall} \& {Cen}}{{Bahcall} \&
  {Cen}}{1992}]{BC1992}
{Bahcall} N.~A.,  {Cen} R.,  1992, ApJ, 398, L81

\bibitem[\protect\citeauthoryear{{Bahcall}, {Fan} \& {Cen}}{{Bahcall}
  et~al.}{1997}]{Bahcall1997}
{Bahcall} N.~A.,  {Fan} X.,    {Cen} R.,  1997, ApJ, 485, L53

\bibitem[\protect\citeauthoryear{{Baum}}{{Baum}}{1962}]{Baum62}
{Baum} W.~A.,  1962, in IAU Symp. 15: Problems of Extra-Galactic Research
  Vol.~15, {Photoelectric Magnitudes and Red-Shifts}.
p.~390

\bibitem[\protect\citeauthoryear{{Bender} et~al.,}{{Bender}
  et~al.}{2001}]{photred00_mod}
{Bender} R.,  et~al., 2001, in Christiani S.,  ed., ESO/ECF/STScI Workshop on
  Deep Fields "{The} {FORS} {Deep} {Field}: {Photometric} {Data} and
  {Photometric} {Redshifts}".
Springer, Berlin, p.~327

\bibitem[\protect\citeauthoryear{{Ben{\'\i}tez}}{{Ben{\'\i}tez}}{2000}]{Benite%
z00}
{Ben{\'\i}tez} N.,  2000, ApJ, 536, 571

\bibitem[\protect\citeauthoryear{{Bode}, {Bahcall}, {Ford} \&
  {Ostriker}}{{Bode} et~al.}{2001}]{Bode2001}
{Bode} P.,  {Bahcall} N.~A.,  {Ford} E.~B.,    {Ostriker} J.~P.,  2001, ApJ,
  551, 15

\bibitem[\protect\citeauthoryear{{Brunner}, {Connolly} \& {Szalay}}{{Brunner}
  et~al.}{1999}]{Brunner1999}
{Brunner} R.~J.,  {Connolly} A.~J.,    {Szalay} A.~S.,  1999, ApJ, 516, 563

\bibitem[\protect\citeauthoryear{{Carrol}, {Press} \& {Turner}}{{Carrol}
  et~al.}{1992}]{CPT:1992}
{Carrol} S.~M.,  {Press} W.~H.,    {Turner} E.~L.,  1992, ARA\&A, 30, 499

\bibitem[\protect\citeauthoryear{{Cole} \& {Lacey}}{{Cole} \&
  {Lacey}}{1996}]{CL1996}
{Cole} S.,  {Lacey} C.,  1996, MNRAS, 281, 716

\bibitem[\protect\citeauthoryear{{Couch}, {Ellis}, {MacLaren} \&
  {Malin}}{{Couch} et~al.}{1991}]{Couch1991}
{Couch} W.~J.,  {Ellis} R.~S.,  {MacLaren} I.,    {Malin} D.~F.,  1991, MNRAS,
  249, 606

\bibitem[\protect\citeauthoryear{{Dalton}, {Maddox}, {Sutherland} \&
  {Efstathiou}}{{Dalton} et~al.}{1997}]{Dalton1997}
{Dalton} G.~B.,  {Maddox} S.~J.,  {Sutherland} W.~J.,    {Efstathiou} G.,
  1997, MNRAS, 289, 263

\bibitem[\protect\citeauthoryear{{Davis}, {Efstathiou}, {Frenk} \&
  {White}}{{Davis} et~al.}{1985}]{DEFW1985}
{Davis} M.,  {Efstathiou} G.,  {Frenk} C.~S.,    {White} S.~D.~M.,  1985, ApJ,
  292, 371

\bibitem[\protect\citeauthoryear{{de Propris}, {Stanford}, {Eisenhardt},
  {Dickinson} \& {Elston}}{{de Propris} et~al.}{1999}]{dePropris1999}
{de Propris} R.,  {Stanford} S.~A.,  {Eisenhardt} P.~R.,  {Dickinson} M.,
  {Elston} R.,  1999, AJ, 118, 719

\bibitem[\protect\citeauthoryear{{Drory}, {Bender}, {Snigula}, {Feulner},
  {Hopp}, {Maraston}, {Hill} \& {de Oliveira}}{{Drory}
  et~al.}{001a}]{Drory2001}
{Drory} N.,  {Bender} R.,  {Snigula} J.,  {Feulner} G.,  {Hopp} U.,  {Maraston}
  C.,  {Hill} G.~J.,    {de Oliveira} C.~M.,  {2001a}, ApJ, 562, L111

\bibitem[\protect\citeauthoryear{{Drory}, {Feulner}, {Bender}, {Botzler},
  {Hopp}, {Maraston}, {Mendes de Oliveira} \& {Snigula}}{{Drory}
  et~al.}{001b}]{MUNICS1}
{Drory} N.,  {Feulner} G.,  {Bender} R.,  {Botzler} C.~S.,  {Hopp} U.,
  {Maraston} C.,  {Mendes de Oliveira} C.,    {Snigula} J.,  {2001b}, MNRAS,
  325, 550

\bibitem[\protect\citeauthoryear{{Efstathiou}, {Frenk}, {White} \&
  {Davis}}{{Efstathiou} et~al.}{1988}]{EFWD1988}
{Efstathiou} G.,  {Frenk} C.~S.,  {White} S.~D.~M.,    {Davis} M.,  1988,
  MNRAS, 235, 715

\bibitem[\protect\citeauthoryear{{Eisenstein} \& {Hut}}{{Eisenstein} \&
  {Hut}}{1998}]{EH1998}
{Eisenstein} D.~J.,  {Hut} P.,  1998, ApJ, 498, 137

\bibitem[\protect\citeauthoryear{{Eke}, {Cole} \& {Frenk}}{{Eke}
  et~al.}{1996}]{Eke1996}
{Eke} V.~R.,  {Cole} S.,    {Frenk} C.~S.,  1996, MNRAS, 282, 263

\bibitem[\protect\citeauthoryear{{Fern{\'a}ndez-Soto}, {Lanzetta} \&
  {Yahil}}{{Fern{\'a}ndez-Soto} et~al.}{1999}]{FLY99}
{Fern{\'a}ndez-Soto} A.,  {Lanzetta} K.~M.,    {Yahil} A.,  1999, ApJ, 513, 34

\bibitem[\protect\citeauthoryear{{Fried}, {von Kuhlmann}, {Meisenheimer},
  {Rix}, {Wolf}, {Hippelein}, {K{\" u}mmel}, {Phleps}, {R{\" o}ser},
  {Thierring} \& {Maier}}{{Fried} et~al.}{2001}]{Fried2001}
{Fried} J.~W.,  {von Kuhlmann} B.,  {Meisenheimer} K.,  {Rix} H.-W.,  {Wolf}
  C.,  {Hippelein} H.~H.,  {K{\" u}mmel} M.,  {Phleps} S.,  {R{\" o}ser} H.~J.,
   {Thierring} I.,    {Maier} C.,  2001, A\&A, 367, 788

\bibitem[\protect\citeauthoryear{{Geller} \& {Huchra}}{{Geller} \&
  {Huchra}}{1983}]{GH1983}
{Geller} M.~J.,  {Huchra} J.~P.,  1983, ApJS, 52, 61

\bibitem[\protect\citeauthoryear{{Giuricin}, {Marinoni}, {Ceriani} \&
  {Pisani}}{{Giuricin} et~al.}{2000}]{GMCP2000}
{Giuricin} G.,  {Marinoni} C.,  {Ceriani} L.,    {Pisani} A.,  2000, ApJ, 543,
  178

\bibitem[\protect\citeauthoryear{{Gladders} \& {Yee}}{{Gladders} \&
  {Yee}}{2000}]{GY2000}
{Gladders} M.~D.,  {Yee} H.~K.~C.,  2000, AJ, 120, 2148

\bibitem[\protect\citeauthoryear{{Goto}, {Sekiguchi}, {Nichol}, {Bahcall},
  {Kim}, {Annis}, {Ivezi{\' c}}, {Brinkmann}, {Hennessy}, {Szokoly} \&
  {Tucker}}{{Goto} et~al.}{2002}]{Goto2002}
{Goto} T.,  {Sekiguchi} M.,  {Nichol} R.~C.,  {Bahcall} N.~A.,  {Kim} R.~S.~J.,
   {Annis} J.,  {Ivezi{\' c}} {\v Z}.,  {Brinkmann} J.,  {Hennessy} G.~S.,
  {Szokoly} G.~P.,    {Tucker} D.~L.,  2002, AJ, 123, 1807

\bibitem[\protect\citeauthoryear{{Gourgoulhon}, {Chamaraux} \&
  {Fouque}}{{Gourgoulhon} et~al.}{1992}]{Gourgoulhon1992}
{Gourgoulhon} E.,  {Chamaraux} P.,    {Fouque} P.,  1992, A\&A, 255, 69

\bibitem[\protect\citeauthoryear{{Heidt} et~al.,}{{Heidt}  et~al.}{2003}]{FDF1}
{Heidt} J.,  et~al., 2003, A\&A, 398, 49

\bibitem[\protect\citeauthoryear{{Hopp} \& {Materne}}{{Hopp} \&
  {Materne}}{1985}]{hopp1985}
{Hopp} U.,  {Materne} J.,  1985, A\&AS, 61, 93

\bibitem[\protect\citeauthoryear{{Huchra}, {Davis}, {Latham} \&
  {Tonry}}{{Huchra} et~al.}{1983}]{Huchra1983}
{Huchra} J.,  {Davis} M.,  {Latham} D.,    {Tonry} J.,  1983, ApJS, 52, 89

\bibitem[\protect\citeauthoryear{{Huchra} \& {Geller}}{{Huchra} \&
  {Geller}}{1982}]{HG1982}
{Huchra} J.~P.,  {Geller} M.~J.,  1982, ApJ, 257, 423

\bibitem[\protect\citeauthoryear{{Jing} \& {Fang}}{{Jing} \&
  {Fang}}{1994}]{JF1994}
{Jing} Y.,  {Fang} L.,  1994, ApJ, 432, 438

\bibitem[\protect\citeauthoryear{{Kepner}, {Fan}, {Bahcall}, {Gunn}, {Lupton}
  \& {Xu}}{{Kepner} et~al.}{1999}]{Kepner1999}
{Kepner} J.,  {Fan} X.,  {Bahcall} N.,  {Gunn} J.,  {Lupton} R.,    {Xu} G.,
  1999, ApJ, 517, 78

\bibitem[\protect\citeauthoryear{{Kim}, {Kepner}, {Postman}, {Strauss},
  {Bahcall}, {Gunn}, {Lupton}, {Annis}, {Nichol}, {Castander}, {Brinkmann},
  {Brunner}, {Connolly}, {Csabai}, {Hindsley}, {Ivezi{\' c}}, {Vogeley} \&
  {York}}{{Kim} et~al.}{2002}]{Kim2002}
{Kim} R.~S.~J.,  {Kepner} J.~V.,  {Postman} M.,  {Strauss} M.~A.,  {Bahcall}
  N.~A.,  {Gunn} J.~E.,  {Lupton} R.~H.,  {Annis} J.,  {Nichol} R.~C.,
  {Castander} F.~J.,  {Brinkmann} J.,  {Brunner} R.~J.,  {Connolly} A.,
  {Csabai} I.,  {Hindsley} R.~B.,  {Ivezi{\' c}} {\v Z}.,  {Vogeley} M.~S.,
  {York} D.~G.,  2002, AJ, 123, 20

\bibitem[\protect\citeauthoryear{{Koo}}{{Koo}}{1985}]{Koo85}
{Koo} D.~C.,  1985, AJ, 90, 418

\bibitem[\protect\citeauthoryear{{Lacey} \& {Cole}}{{Lacey} \&
  {Cole}}{1994}]{LC1994}
{Lacey} C.,  {Cole} S.,  1994, MNRAS, 271, 676

\bibitem[\protect\citeauthoryear{{Lidman} \& {Peterson}}{{Lidman} \&
  {Peterson}}{1996}]{LP1996}
{Lidman} C.~E.,  {Peterson} B.~A.,  1996, AJ, 112, 2454

\bibitem[\protect\citeauthoryear{{Lin}, {Kirshner}, {Shectman}, {Landy},
  {Oemler}, {Tucker} \& {Schechter}}{{Lin} et~al.}{1996}]{LCRS96a}
{Lin} H.,  {Kirshner} R.~P.,  {Shectman} S.~A.,  {Landy} S.~D.,  {Oemler} A.,
  {Tucker} D.~L.,    {Schechter} P.~L.,  1996, ApJ, 464, 60

\bibitem[\protect\citeauthoryear{{Lineweaver}, {Tenorio}, {Smoot}, {Keegstra},
  {Banday} \& {Lubin}}{{Lineweaver} et~al.}{1996}]{Lineweaver1996}
{Lineweaver} C.~H.,  {Tenorio} L.,  {Smoot} G.~F.,  {Keegstra} P.,  {Banday}
  A.~J.,    {Lubin} P.,  1996, ApJ, 470, 38

\bibitem[\protect\citeauthoryear{{Lobo}, {Iovino}, {Lazzati} \&
  {Chincarini}}{{Lobo} et~al.}{2000}]{Lobo2000}
{Lobo} C.,  {Iovino} A.,  {Lazzati} D.,    {Chincarini} G.,  2000, A\&A, 360,
  896

\bibitem[\protect\citeauthoryear{{Marzke}, {McCarthy}, {Persson}, {Oemler},
  {Dressler}, {Yan}, {Carlberg}, {Abraham}, {Ellis}, {Firth}, {Mackay} \&
  {McMahon}}{{Marzke} et~al.}{1999}]{Marzke1999}
{Marzke} R.,  {McCarthy} P.~J.,  {Persson} E.,  {Oemler} A.,  {Dressler} A.,
  {Yan} M.~B.~L.,  {Carlberg} R.,  {Abraham} R.,  {Ellis} R.,  {Firth} A.,
  {Mackay} C.,    {McMahon} R.~G.,  1999, in ASP Conf. Ser. 191: Photometric
  Redshifts and the Detection of High Redshift Galaxies {The Las Campanas
  Infrared Survey}.
p.~148

\bibitem[\protect\citeauthoryear{{Materne}}{{Materne}}{1978}]{Materne1978}
{Materne} J.,  1978, A\&A, 63, 401

\bibitem[\protect\citeauthoryear{{McCarthy}, {Carlberg}, {Marzke}, {Chen},
  {Firth}, {McMahon}, {Wilson}, {Persson}, {Ellis}, {Abraham}, {Lahav},
  {Oemler}, {Sabbey} \& {Somerville}}{{McCarthy} et~al.}{001b}]{McC012}
{McCarthy} P.,  {Carlberg} R.,  {Marzke} R.,  {Chen} H.,  {Firth} A.,
  {McMahon} R.,  {Wilson} J.,  {Persson} E.,  {Ellis} R.,  {Abraham} R.,
  {Lahav} O.,  {Oemler} A.,  {Sabbey} C.,    {Somerville} R.,  {2001b}, in Deep
  Fields {Clustering of Very Red Galaxies in the Las Campanas IR Survey}.
p.~247

\bibitem[\protect\citeauthoryear{{McCarthy}, {Carlberg}, {Chen}, {Marzke},
  {Firth}, {Ellis}, {Persson}, {McMahon}, {Lahav}, {Wilson}, {Martini},
  {Abraham}, {Sabbey}, {Oemler}, {Murphy}, {Somerville}, {Beckett}, {Lewis} \&
  {MacKay}}{{McCarthy} et~al.}{001a}]{McC011}
{McCarthy} P.~J.,  {Carlberg} R.~G.,  {Chen} H.-W.,  {Marzke} R.~O.,  {Firth}
  A.~E.,  {Ellis} R.~S.,  {Persson} S.~E.,  {McMahon} R.~G.,  {Lahav} O.,
  {Wilson} J.,  {Martini} P.,  {Abraham} R.~G.,  {Sabbey} C.~N.,  {Oemler} A.,
  {Murphy} D.~C.,  {Somerville} R.~S.,  {Beckett} M.~G.,  {Lewis} J.~R.,
  {MacKay} C.~D.,  {2001a}, ApJ, 560, L131

\bibitem[\protect\citeauthoryear{{Merch{\' a}n}, {Maia} \& {Lambas}}{{Merch{\'
  a}n} et~al.}{2000}]{MML2000}
{Merch{\' a}n} M.~E.,  {Maia} M.~A.~G.,    {Lambas} D.~G.,  2000, ApJ, 545, 26

\bibitem[\protect\citeauthoryear{{Nilson}}{{Nilson}}{1973}]{Nilson1973}
{Nilson} P.,  1973, {Uppsala General Catalogue of Galaxies}.
Acta Universitatis Upsaliensis.~Nova Acta Regiae Societatis Scientiarum
  Upsaliensis - Uppsala Astronomiska Observatoriums Annaler, Uppsala:
  Astronomiska Observatorium, 1973

\bibitem[\protect\citeauthoryear{{Paredes}, {Jones} \& {Martinez}}{{Paredes}
  et~al.}{1995}]{PJM1995}
{Paredes} S.,  {Jones} B.~J.~T.,    {Martinez} V.~J.,  1995, MNRAS, 276, 1116

\bibitem[\protect\citeauthoryear{{Peebles}}{{Peebles}}{1993}]{Peebles1993}
{Peebles} P.~J.~E.,  1993, {Principles of Physical Cosmology}.
Princeton Series in Physics, Princeton, NJ: Princeton University Press

\bibitem[\protect\citeauthoryear{{Plionis}, {Barrow} \& {Frenk}}{{Plionis}
  et~al.}{1991}]{Plionis1991}
{Plionis} M.,  {Barrow} J.~D.,    {Frenk} C.~S.,  1991, MNRAS, 249, 662

\bibitem[\protect\citeauthoryear{{Postman}, {Lubin}, {Gunn}, {Oke}, {Hoessel},
  {Schneider} \& {Christensen}}{{Postman} et~al.}{1996}]{Postman1996}
{Postman} M.,  {Lubin} L.~M.,  {Gunn} J.~E.,  {Oke} J.~B.,  {Hoessel} J.~G.,
  {Schneider} D.~P.,    {Christensen} J.~A.,  1996, AJ, 111, 615

\bibitem[\protect\citeauthoryear{{Postman}, {Lubin} \& {Oke}}{{Postman}
  et~al.}{2001}]{Postman2001}
{Postman} M.,  {Lubin} L.~M.,    {Oke} J.~B.,  2001, AJ, 122, 1125

\bibitem[\protect\citeauthoryear{{Press} \& {Schechter}}{{Press} \&
  {Schechter}}{1974}]{PS1974}
{Press} W.~H.,  {Schechter} P.,  1974, ApJ, 187, 425

\bibitem[\protect\citeauthoryear{{Ramella}, {Boschin}, {Fadda} \&
  {Nonino}}{{Ramella} et~al.}{2001}]{Ramella2001}
{Ramella} M.,  {Boschin} W.,  {Fadda} D.,    {Nonino} M.,  2001, A\&A, 368, 776

\bibitem[\protect\citeauthoryear{{Ramella}, {Geller} \& {Huchra}}{{Ramella}
  et~al.}{1989}]{RGH1989}
{Ramella} M.,  {Geller} M.~J.,    {Huchra} J.~P.,  1989, ApJ, 344, 57

\bibitem[\protect\citeauthoryear{{Ramella}, {Geller}, {Pisani} \& {da
  Costa}}{{Ramella} et~al.}{2002}]{RGPdC2002}
{Ramella} M.,  {Geller} M.~J.,  {Pisani} A.,    {da Costa} L.~N.,  2002, AJ,
  123, 2976

\bibitem[\protect\citeauthoryear{{Ramella}, {Pisani} \& {Geller}}{{Ramella}
  et~al.}{1997}]{RPG1997}
{Ramella} M.,  {Pisani} A.,    {Geller} M.~J.,  1997, AJ, 113, 483

\bibitem[\protect\citeauthoryear{{Schuecker} \& {Boehringer}}{{Schuecker} \&
  {Boehringer}}{1998}]{SB1998}
{Schuecker} P.,  {Boehringer} H.,  1998, A\&A, 339, 315

\bibitem[\protect\citeauthoryear{{Shectman}, {Landy}, {Oemler}, {Tucker},
  {Lin}, {Kirshner} \& {Schechter}}{{Shectman} et~al.}{1996}]{Shectman1996}
{Shectman} S.~A.,  {Landy} S.~D.,  {Oemler} A.,  {Tucker} D.~L.,  {Lin} H.,
  {Kirshner} R.~P.,    {Schechter} P.~L.,  1996, ApJ, 470, 172

\bibitem[\protect\citeauthoryear{{Slezak}, {Bijaoui} \& {Mars}}{{Slezak}
  et~al.}{1990}]{SBM1990}
{Slezak} E.,  {Bijaoui} A.,    {Mars} G.,  1990, A\&A, 227, 301

\bibitem[\protect\citeauthoryear{{Trasarti-Battistoni}}{{Trasarti-Battistoni}}%
{1998}]{TB1998}
{Trasarti-Battistoni} R.,  1998, A\&AS, 130, 341

\bibitem[\protect\citeauthoryear{{Trentham}}{{Trentham}}{1998}]{Trentham1998}
{Trentham} N.,  1998, MNRAS, 294, 193

\bibitem[\protect\citeauthoryear{{Tucker}, {Oemler}, {Kirshner}, {Lin},
  {Shectman}, {Landy}, {Schechter}, {Muller}, {Gottlober} \&
  {Einasto}}{{Tucker} et~al.}{1997}]{Tucker1997}
{Tucker} D.~L.,  {Oemler} A.,  {Kirshner} R.~P.,  {Lin} H.,  {Shectman} S.~A.,
  {Landy} S.~D.,  {Schechter} P.~L.,  {Muller} V.,  {Gottlober} S.,
  {Einasto} J.,  1997, MNRAS, 285, L5

\bibitem[\protect\citeauthoryear{{Tucker}, {Oemler}, {Hashimoto}, {Shectman},
  {Kirshner}, {Lin}, {Landy}, {Schechter} \& {Allam}}{{Tucker}
  et~al.}{2000}]{Tucker2000}
{Tucker} D.~L.,  {Oemler} A.~J.,  {Hashimoto} Y.,  {Shectman} S.~A.,
  {Kirshner} R.~P.,  {Lin} H.,  {Landy} S.~D.,  {Schechter} P.~L.,    {Allam}
  S.~S.,  2000, ApJS, 130, 237

\bibitem[\protect\citeauthoryear{{Tully}}{{Tully}}{1980}]{T1980}
{Tully} R.~B.,  1980, ApJ, 237, 390

\bibitem[\protect\citeauthoryear{{Tully}}{{Tully}}{1987}]{Tully1987}
{Tully} R.~B.,  1987, ApJ, 321, 280

\bibitem[\protect\citeauthoryear{{Turner} \& {Gott}}{{Turner} \&
  {Gott}}{1976}]{TG1976}
{Turner} E.~L.,  {Gott} J.~R.,  1976, ApJS, 32, 409

\bibitem[\protect\citeauthoryear{{Ueda}, {Itoh} \& {Suto}}{{Ueda}
  et~al.}{1993}]{Ueda1993}
{Ueda} H.,  {Itoh} M.,    {Suto} Y.,  1993, ApJ, 408, 3

\bibitem[\protect\citeauthoryear{{Valageas}, {Lacey} \& {Schaeffer}}{{Valageas}
  et~al.}{2000}]{VLS2000}
{Valageas} P.,  {Lacey} C.,    {Schaeffer} R.,  2000, MNRAS, 311, 234

\bibitem[\protect\citeauthoryear{{Williams}, {Blacker}, {Dickinson}, {Dixon},
  {Ferguson}, {Fruchter}, {Giavalisco}, {Gilliland}, {Heyer}, {Katsanis},
  {Levay}, {Lucas}, {McElroy}, {Petro}, {Postman}, {Adorf} \&
  {Hook}}{{Williams} et~al.}{1996}]{HDF96}
{Williams} R.~E.,  {Blacker} B.,  {Dickinson} M.,  {Dixon} W. V.~D.,
  {Ferguson} H.~C.,  {Fruchter} A.~S.,  {Giavalisco} M.,  {Gilliland} R.~L.,
  {Heyer} I.,  {Katsanis} R.,  {Levay} Z.,  {Lucas} R.~A.,  {McElroy} D.~B.,
  {Petro} L.,  {Postman} M.,  {Adorf} H.,    {Hook} R.,  1996, AJ, 112, 1335

\bibitem[\protect\citeauthoryear{{Williams} et~al.,}{{Williams}
  et~al.}{2000}]{HDFS00}
{Williams} R.~E.,  et~al., 2000, AJ, 120, 2735

\bibitem[\protect\citeauthoryear{{Wolf}, {Dye}, {Kleinheinrich},
  {Meisenheimer}, {Rix} \& {Wisotzki}}{{Wolf} et~al.}{001b}]{combo17}
{Wolf} C.,  {Dye} S.,  {Kleinheinrich} M.,  {Meisenheimer} K.,  {Rix} H.-W.,
  {Wisotzki} L.,  {2001b}, A\&A, 377, 442

\bibitem[\protect\citeauthoryear{{Wolf}, {Meisenheimer}, {R{\" o}ser},
  {Beckwith}, {Chaffee}, {Fried}, {Hippelein}, {Huang}, {K{\" u}mmel}, {von
  Kuhlmann}, {Maier}, {Phleps}, {Rix}, {Thommes} \& {Thompson}}{{Wolf}
  et~al.}{001a}]{CADIS1}
{Wolf} C.,  {Meisenheimer} K.,  {R{\" o}ser} H.-J.,  {Beckwith} S.~V.~W.,
  {Chaffee} F.~H.,  {Fried} J.,  {Hippelein} H.,  {Huang} J.-S.,  {K{\" u}mmel}
  M.,  {von Kuhlmann} B.,  {Maier} C.,  {Phleps} S.,  {Rix} H.-W.,  {Thommes}
  E.,    {Thompson} D.,  {2001a}, A\&A, 365, 681

\bibitem[\protect\citeauthoryear{{Zwicky}, {Herzog} \& {Wild}}{{Zwicky}
  et~al.}{1968}]{Zwicky6168}
{Zwicky} F.,  {Herzog} E.,    {Wild} P.,  1961-1968, {Catalogue of Galaxies and
  of Clusters of Galaxies, 6 volumes}.
Pasadena: California Institute of Technology (CIT), 1961-1968

\bibitem[\protect\citeauthoryear{{Zwicky} \& {Zwicky}}{{Zwicky} \&
  {Zwicky}}{1971}]{Zwicky1971}
{Zwicky} F.,  {Zwicky} M.~A.,  1971, {Catalogue of Selected Compact Galaxies
  and of Post-Eruptive Galaxies}.
Guemligen

\end{thebibliography}



\label{lastpage}

\end{document}